\documentclass[letterpaper,11pt]{article}
\pdfoutput=1

\usepackage{xspace}
\usepackage{slashed}
\usepackage{array,multirow}
\usepackage{graphicx}
\usepackage{graphics}
\usepackage{epsfig}
\usepackage{amsfonts}
\usepackage{amsmath}
\usepackage{amssymb}
\usepackage{dsfont}
\usepackage{color}
\usepackage{xcolor}
\usepackage{cite}
\usepackage{pdflscape}
\usepackage{pifont}
\usepackage{pgfplots}
\usepackage{tikz} 
\usepackage{pgfplotstable}
\usepackage{enumitem}

\usepackage{bm}  %

\newcommand{\be}{\begin{equation}}
\newcommand{\ee}{\end{equation}}
\newcommand{\bea}{\begin{eqnarray}}
\newcommand{\eea}{\end{eqnarray}}
\newcommand{\ba}{\begin{array}}
\newcommand{\ea}{\end{array}}

\renewcommand\({\left(}
\renewcommand\){\right)}
\renewcommand\[{\left[}
\renewcommand\]{\right]}

\newcommand{\dd}{{\rm d}}
\newcommand{\e}{{\rm e}}

\newcommand{\eref}[1]{Eq.~(\ref{#1})}

\def\E{\mathcal{E}}
\def\O{\mathcal{O}}

\def\A{\mathcal{A}}
\def\nn{\nonumber}

\newcommand{\dbar}{ {\slashed{\rm d}}}

\newcommand{\CPV}{CP\!\!\!\!\!\!\!\!\raisebox{0pt}{\small$\diagup$}}
\renewcommand{\Re}{{\rm Re}}
\renewcommand{\Im}{{\rm Im}}

\DeclareMathOperator{\Tr}{Tr}

\def\slashchar#1{\setbox0=\hbox{$#1$}           
  \dimen0=\wd0                                    
  \setbox1=\hbox{/} \dimen1=\wd1                  
  \ifdim\dimen0>\dimen1                           
    \rlap{\hbox to \dimen0{\hfil/\hfil}}            
    #1                                             
  \else                                          
    \rlap{\hbox to \dimen1{\hfil$#1$\hfil}}        
    /                                           
 \fi}

\usepackage{tikz}
\usetikzlibrary{arrows,shapes}
\usetikzlibrary{trees}
\usetikzlibrary{matrix,arrows} 				
\usetikzlibrary{positioning}				
\usetikzlibrary{calc,through}				
\usetikzlibrary{decorations.pathreplacing}  
\usepackage{pgffor}							

\usetikzlibrary{decorations.pathmorphing}	
\usetikzlibrary{decorations.markings}
\tikzset{
    vector/.style={decorate, decoration={snake}, draw},
	provector/.style={decorate, decoration={snake,amplitude=2.5pt}, draw},
	antivector/.style={decorate, decoration={snake,amplitude=-2.5pt}, draw},
    fermion/.style={draw=black, postaction={decorate},
        decoration={markings,mark=at position .55 with {\arrow[draw=black]{>}}}},
    fermionbar/.style={draw=black, postaction={decorate},
        decoration={markings,mark=at position .55 with {\arrow[draw=black]{<}}}},
    fermionnoarrow/.style={draw=black},
    gluon/.style={decorate, draw=black,
        decoration={coil,amplitude=4pt, segment length=5pt}},
    scalar/.style={dashed,draw=black, postaction={decorate},
        decoration={markings,mark=at position .55 with {\arrow[draw=black]{>}}}},
    scalarbar/.style={dashed,draw=black, postaction={decorate},
        decoration={markings,mark=at position .55 with {\arrow[draw=black]{<}}}},
    scalarnoarrow/.style={dashed,draw=black},
    electron/.style={draw=black, postaction={decorate},
        decoration={markings,mark=at position .55 with {\arrow[draw=black]{>}}}},
	bigvector/.style={decorate, decoration={snake,amplitude=4pt}, draw},
}

\begin{document}

\begin{titlepage}

\begin{flushright}
Nikhef-2017-044
\end{flushright}

\vspace{2.0cm}

\begin{center}
{\LARGE  \bf 
Electroweak Baryogenesis and the \\ \vspace{3mm}
Standard Model Effective Field Theory

}
\vspace{2.4cm}

{\large \bf  Jordy de Vries$^a$, Marieke Postma$^a$, Jorinde van de Vis$^a$, Graham White$^{b,c}$ } 
\vspace{0.5cm}

\vspace{0.25cm}

{\large 
$^a$ 
{\it Nikhef, Theory Group, Science Park 105, 1098 XG, Amsterdam, The Netherlands}}

\vspace{0.25cm}
{\large 
$^b$ 
{\it 
ARC Centre of Excellence for Particle Physics at the Terascale
School of Physics and Astronomy, Monash University, Victoria 3800, Australia
}}

\vspace{0.25cm}
{\large 
$^c$ 
{\it 
TRIUMF, 4004 Wesbrook Mall, Vancouver, British Columbia V6T 2A3, Canada
}}

\end{center}

\vspace{1.5cm}

\begin{abstract}

We investigate electroweak baryogenesis within the framework of the Standard Model Effective Field Theory. The Standard Model Lagrangian is supplemented by dimension-six operators that facilitate a strong first-order electroweak phase transition and provide sufficient CP violation.  Two explicit scenarios are studied that are related via the classical equations of motion and are therefore identical at leading order in the effective field theory expansion. We demonstrate that formally higher-order dimension-eight corrections lead to large modifications of the matter-antimatter asymmetry. The effective field theory expansion breaks down in the modified Higgs sector due to the requirement of a first-order phase transition.  We investigate the source of the breakdown in detail and show how it is transferred to the CP-violating sector.  We briefly discuss possible modifications of the effective field theory framework.

\end{abstract}

\vfill
\end{titlepage}


\section{Introduction}\label{sec:intro}

The asymmetry between baryons and anti-baryons, characterized by the ratio of densities of baryon number and entropy, has been determined by two independent methods \cite{Cooke:2013cba,Ade:2015xua}
\begin{equation}
  Y_B = \frac{n_B}{s} = \begin{cases}
    8.2\text{-}9.4 \times 10^{-11}  & {\rm Big\,Bang\,Nucleosynthesis}  \\ 
    8.65 \pm 0.09 \times 10^{-11} & \rm{PLANCK}~  
  \end{cases}\label{BAUExp} 
\end{equation}
which are in good agreement. The nonzero value of $Y_B$ provides one of the strongest indications that the Standard Model (SM) of particle physics is incomplete. While the SM has a sufficiently rich structure to in principle fulfill the three Sakharov conditions \cite{Sakharov:1967dj}, in practice it gives rise to an asymmetry that is too small by many orders of magnitude. The first problem is that the electroweak phase transition (EWPT) is a cross-over transition, whereas the required strong first-order transition can only occur for a much lighter Higgs boson than is observed \cite{Gurtler:1997hr,Laine:1998jb,Csikor:1998eu,Aoki:1999fi}. The second problem is that the amount of CP violation in the SM is  not sufficient to produce the observed baryon asymmetry \cite{Gavela:1993ts,Huet:1994jb,Gavela:1994dt}.

Understanding why baryons are more abundant than anti-baryons thus requires beyond-the-SM (BSM) physics. Such BSM physics could live at a very high energy scale, decoupled from the electroweak scale, as occurs for instance in (most) scenarios of leptogenesis. Such  scenarios, while well motivated, will be difficult to probe in current and upcoming experiments although measurements of neutrinoless double beta decay would point towards them. In scenarios of electroweak baryogenesis (EWBG) \cite{Trodden:1998ym,Morrissey:2012db,White:2016nbo}, however, the scale of BSM physics cannot be much higher than the electroweak scale which makes the scenario more testable. In particular, searches for new scalars, precision measurements of Higgs couplings, and electric dipole moment (EDM) experiments all probe different aspects of EWBG scenarios.  

The above considerations have led to a large number of SM extensions that can lead to successful EWBG. Depending on the BSM details, such as the particle content and symmetries, different tests are required and each scenario requires a detailed phenomenological study. It would be a great advantage if the crucial aspects of all these models can be tested in a single framework. In principle, the SM Effective Field Theory (SM-EFT) could provide such a framework \cite{Grojean:2004xa,Bodeker:2004ws,Huber:2006ri,Delaunay:2007wb,Grinstein:2008qi,
  Damgaard:2015con, Kobakhidze:2015xlz, Balazs:2016yvi} as it provides a model-independent parametrization of BSM physics.   
  The SM-EFT assumes that any BSM degrees of freedom are sufficiently heavy, such that they can be integrated out and that their low-energy effects can be captured by effective gauge-invariant operators containing just SM degrees of freedom.  While an infinite number of effective operators exist, they can be organized by their dimension. The higher the dimension of the operators, the more suppressed their low-energy effects are by powers of $E/\Lambda$, where $E$ is a typical low-energy scale, such as the electroweak scale, and $\Lambda$ the scale of BSM physics.  The first operators relevant for EWBG appear at dimension-six. If the SM-EFT is suitable for the description of EWBG, it would provide an attractive framework as the dimension-six operators have to a large extent been connected to low- and high-energy experiments  already, while the EFT operators can be easily matched to specific UV-complete models.  

The applicability of the SM-EFT requires a perturbative expansion in $E/\Lambda$, which is potentially dangerous 
for EWBG applications. Extending the SM scalar potential with a dimension-six cubic interaction to ensure a strong first-order EWPT requires a relatively low scale $\Lambda \lesssim 800$ GeV \cite{Grojean:2004xa}, which can lead to a mismatch between calculations in the SM-EFT and specific UV-complete models, see for instance Ref.~\cite{Damgaard:2015con} for an analysis of the singlet-extended SM. Furthermore, EDM constraints on dimension-six CP-violating (CPV) operators potentially relevant for EWBG are typically in the multi-TeV range \cite{Brod:2013cka, Chien:2015xha, Cirigliano:2016njn,Fuyuto:2017xup}. This difference in scale can be accommodated by assuming a different threshold for the CPV dimension-six operators such that $\Lambda_{CP} > \Lambda$ \cite{Huber:2006ri}. In this way, it might be possible to use EFT techniques for the CPV sector despite the relatively low scale required for a strong first-order EWPT.

In this work we investigate a related issue of the EFT approach to EWBG. As mentioned, the EFT approach requires that the effects of higher-dimensional operators are suppressed with respect to the lower-dimensional ones. For energies ($E$) around the electroweak scale and $\Lambda \simeq 800$ GeV, the expansion parameter $(E/\Lambda)^2$ seems at first sight to be sufficiently small for a perturbative expansion. 
In practice, the necessity of a first-order phase transition requires a fine balance between dimension-two, -four, and -six contributions to the Higgs potential. While no such balance is necessary for the CP-violating sector, successful EWBG requires an interplay of the scalar and CPV sectors, such that formally higher-order corrections to the latter might become relevant as well. To study this, we consider two specific EFTs which can be related via the classical equations of motion (EOMs). EOMs can be applied to EFTs to reduce the number of operators in the EFT basis \cite{Arzt:1993gz}. Operators related via EOMs lead to identical observables up to higher-order corrections in the EFT expansion. That is, if the EFT is working satisfactory the two EFTs under investigation should lead to the same baryon asymmetry modulo small corrections. The main goal of our work is to perform a detailed test of this hypothesis. 

A somewhat similar study was performed in Ref.~\cite{Balazs:2016yvi}, where it was concluded that the derivative operators in the EFT can no longer be eliminated by EOMs without explicitly specifying the dynamics of the phase transition.  We improve on these results by carefully investigating --- both analytically and numerically --- the redundancy of the operators in the EFT, including important thermal effects.  We also improve the EDM phenomenology with respect to Ref.~\cite{Balazs:2016yvi}, which neglected several relevant contributions.

Our study allows us to pin down where and how the EFT approach breaks down for the application of EWBG. We find that scenarios that are identical up to higher-order dimension-eight corrections lead to large differences in the baryon asymmetry. The breakdown of the EFT is not specific to the EWBG calculation and in principle also arises at zero temperature where certain CPV interactions get $\mathcal O(1)$-corrections from dimension-eight operators. However, these interactions are largely unconstrained, and as far as the EDM phenomenology is concerned the scenarios that are related by the EOMs are equivalent. In the context of EWBG, however, we find that dimension-eight corrections strongly modify the strength of the CPV source term that drives the creation of the matter-antimatter asymmetry. While this modification is partially washed out due to SM processes that are active during the phase transition, it still leads to a reduction of the matter-antimatter asymmetry by a factor $\mathcal O(4)$. Higher-dimensional CPV operators can therefore not be neglected.


Our paper is organized as follows. In Sect.~\ref{EFT} we introduce the SM-EFT operators we consider and how they are related via the EOMs. We also obtain the EDM constraints on the CPV operators. In Sect.~\ref{bubble} we discuss details of the EWPT. In Sect.~\ref{asymmetry} we review the derivation of the transport equations that describe the plasma in front of the bubble walls.  We focus on how the source term that drives the asymmetry depends on the CPV operators.  It is important to take thermal corrections to the CPV operators into account and these are calculated in Appendix~\ref{A:thermal}.  The baryon asymmetry is calculated in Sect. \ref{discussion}.  Most details of the solution of the transport equations and the values of the parameters that are used in the computation are delegated to Appendix~\ref{A:input}.  With the calculated asymmetries we test the impact of formally higher-order corrections, and identify the source for the breakdown of the EFT expansion.  We summarize, conclude, and give an outlook in Sect.~\ref{conclusions}.


\section{Effective scenarios for electroweak baryogenesis}\label{EFT} 
We begin by defining the SM Lagrangian. We write the Lagrangian in terms of left-handed quark and lepton doublets, $q_L$, and, $l_L$, respectively, and right-handed singlets $u_R$, $d_R$, and $e_R$. The field $\varphi$ represents the $SU_L(2)$ Higgs doublet of scalar fields 
$\varphi^a$. We define
$\tilde \varphi^a = \epsilon^{ab} \varphi^{b*}$,
where $\epsilon^{ab}$ is the antisymmetric tensor in two dimensions
($\epsilon^{12}=+1$). The covariant derivative is given by 
\begin{eqnarray}\label{Cov}
D_{\mu} = \partial_{\mu}  - i \frac{g_s}{2}\, G^a_{\mu} \lambda^a 
- i \frac{g}{2}\, W^i_{\mu} \tau^i - i g^{\prime} Y B_{\mu}\,,
\end{eqnarray}
where $g_s$,
$g$, and $g^{\prime}$ are, respectively, the $SU_c(3)$, $SU_L(2)$, and $U_Y(1)$
coupling constants. $\lambda^a/2$ and $\tau^i/2$ denote $SU(3)$ and $SU(2)$ generators, 
in the representation of the field on which the derivative acts. The hypercharge assignments, $Y$, are 
$1/6$, $2/3$, $-1/3$, $-1/2$, $-1$, and $1/2$ for $q_L$, $u_R$, $d_R$, $l_L$, $e_R$, and $\varphi$, respectively. The 
field strengths are
\begin{eqnarray}
G^a_{\mu \nu} &=& \partial_{\mu} G^a_{\nu} - \partial_{\nu} G^a_{\mu}
- g_s f^{a b c} G^b_{\mu} G^c_{\nu}\,, \\
W^i_{\mu \nu} &=& \partial_{\mu} W^i_{\nu} - \partial_{\nu} W^i_{\mu} -
g \epsilon^{i j k} W^j_{\mu} W^k_{\nu}\,, \\
B_{\mu \nu} &=& \partial_{\mu} B_{\nu} - \partial_{\nu} B_{\mu}\,,
\end{eqnarray}
with
$f^{abc}$ and $\epsilon^{ijk}$ denoting 
the $SU(3)$ and $SU(2)$ structure constants. The SM Lagrangian is then written as
\begin{eqnarray}
\mathcal L_{SM}  &=&  
-\frac{1}{4} 
\left(G^a_{\mu \nu} G^{a\, \mu \nu}
+W^{i}_{\mu\nu} W^{i\, \mu \nu}
+B_{\mu \nu} B^{\mu \nu}\right)
\nonumber \\ 
& &
+ \bar q_L i \slashchar{D}\, q_L + \bar u_R i \slashchar{D}\, u_R 
+ \bar d_R i \slashchar{D}\, d_R
+ \bar l_L i \slashchar D\, l_L + \bar e_R i \slashchar D\, e_R 
+  (D_{\mu} \varphi)^{\dagger} D^{\mu} \varphi 
\nonumber \\  
& & - \mu^2 \varphi^{\dagger} \varphi -
\lambda (\varphi^{\dagger} \varphi)^2 
- \left( \bar q_L Y_u \tilde \varphi\,u_R +  \bar q_L Y_d \varphi \,d_R  
+ \bar l_L Y_e \varphi \,e_R  + \textrm{h.c.}\right)\,. \label{SM}
\end{eqnarray}
We have suppressed fermion generation indices, but note that the Yukawa matrices are general $3\times 3$ matrices in flavor space. In this work, we are mainly interested in interactions of the third generation of quarks.  
We neglect the Yukawa couplings to light fermions, but make an exception for the electron Yukawa which plays an important role when considering EDM constraints. We have left out the topological theta terms which play no role in our discussion. 

The full set of dimension-six gauge-invariant operators was constructed in Ref.~\cite{Buchmuller:1985jz} and updated in Ref.~\cite{Grzadkowski:2010es}. There exist a large set of operators but only relatively few have impact on EWBG \cite{Grinstein:2008qi,Balazs:2016yvi}. Here, we consider two specific scenarios, which we label by scenario \textbf{A} and \textbf{B}, in which we consider a small subset of dimension-six operators:
\begin{enumerate}[label=\bf{\Alph*}]
\item Here we extend the SM Lagrangian by two dimension-six operators
\begin{eqnarray}\label{dim6A}
\mathcal L^{(A)}_{6}&=&- \kappa (\varphi^\dagger \varphi)^3 - \left[ C_Y\, \bar Q_L  y_t \tilde \varphi \,t_R\,(\varphi^\dagger \varphi)+ \textrm{h.c.}\right]\,,
\end{eqnarray}
where $\kappa \sim \Lambda^{-2}$ and $C_Y \sim \Lambda^{-2}_{CP}$ are dimension-six couplings. $Q_L$ and $y_t$ denote, respectively,  the left-handed doublet of the third-generation quarks and the $(33)$-component of the up-type Yukawa-coupling matrix. The first term in Eq.~\eqref{dim6A} modifies the scalar potential and will be used to ensure a strong first-order EWPT. The second term is a dimension-six modification of the top Yukawa coupling which causes a misalignment between the top-quark mass and the top-Higgs coupling such that the latter can obtain a physical CPV phase. In fact, for simplicity we consider a purely imaginary coupling $C_Y = i \tilde c_Y$, with $\tilde c_Y^* = \tilde c_Y$. This particular choice of dimension-six operators has been well studied  \cite{Bodeker:2004ws,Huber:2006ri,Kobakhidze:2015xlz, Balazs:2016yvi} and is sometimes called the minimal EWBG scenario \cite{Huber:2006ri}. 

\item In this scenario we add the same modification to the scalar potential, but consider a different CPV structure. We use
\begin{eqnarray}\label{dim6B}
\mathcal L^{(B)}_{6}&=&- \kappa (\varphi^\dagger \varphi)^3 - \alpha \left[ C_{DD}\, \bar Q_L  D^2\tilde \varphi \,t_R\, + C_{DD}\left(\bar Q^a_L t_R\right)\,\epsilon^{ab}\,\left(\bar e_L^{b}  y_e e_R\right)+ \textrm{h.c.}\right]\,,
\end{eqnarray}
where $e_L$ and $y_e$ denote, respectively, the lepton doublet of the first generation and the real electron Yukawa coupling. $\alpha$ is a real constant introduced for normalization purposes. The second term provides the dimension-six CPV source for EWBG, while the third term describes a CPV top-electron coupling and is introduced for later convenience. As in scenario \textbf{A} we consider a purely imaginary coupling $C_{DD} = i \tilde c_{DD}$, with $\tilde c_{DD}^* = \tilde c_{DD}$. 
\end{enumerate}

It is possible to relate the two scenarios via the classical EOM for the scalar field \cite{Arzt:1993gz}. From the Euler-Lagrange equations we obtain
\begin{eqnarray}
(D^2 \varphi^* )^a 
&=&- \mu^2 (\varphi^*)^a - 2 \lambda (\varphi^\dagger \varphi) (\varphi^*)^a - 3 \kappa (\varphi^\dagger \varphi)^2 (\varphi^*)^a + \epsilon^{ab} \bar t_R y_t Q^b_L  - \bar e_L^a y_e e_R \,,
\end{eqnarray}
where we neglected the Yukawa couplings to other fermions and a term proportional to $ \Lambda^{-2}_{CP}$. 
Applying the EOM to Eq.~\eqref{dim6B} shifts the Lagrangian into\footnote{Here we used that $C_{DD}\,(\bar Q_L t_R)(\bar t_R Q_L) + \mathrm{h.c.} =0$, for purely imaginary $C_{DD}$.}
\be\label{dim6eom}
\mathcal L^{(B)}_{6} \rightarrow \mathcal L^{(\mathrm{EOM})}_{6}=- \kappa (\varphi^\dagger \varphi)^3 + \alpha \left[ \mu^2 C_{DD}\,\bar Q_L\,\tilde \varphi\, t_R + 2 \lambda C_{DD}\,\bar Q_L\,\tilde \varphi\, t_R\,(\varphi^\dagger \varphi)   + C_8 O_8 \right]\,,
\ee
where the top-electron term in Eq.~\eqref{dim6B} has cancelled and the dimension-eight piece is given by
\begin{equation}
C_8 O_8 = 3\,  \kappa\, C_{DD}\,\bar Q_L\,\tilde \varphi\, t_R\,(\varphi^\dagger \varphi)^2\, ,
\end{equation}
which scales as $\sim \Lambda^{-2}\Lambda_{CP}^{-2}$\,. If the EFT is working satisfactory this term should give rise to small corrections compared to the dimension-six terms in Eq.~\eqref{dim6eom}. It is possible to simplify Eq.~\eqref{dim6eom} by redefining the $Q_L$ and $t_R$ in order to absorb the $\mu^2 C_{DD}$ term into the SM top-Yukawa coupling. The resulting Lagrangian then becomes
\begin{eqnarray}\label{dim6eom2}
 \mathcal L^{(\mathrm{EOM})}_{6}&=&- \kappa (\varphi^\dagger \varphi)^3 + \alpha \left[  2 \lambda C_{DD}\,\bar Q_L\,\tilde \varphi\, t_R\,(\varphi^\dagger \varphi)   + C_8 O_8 \right]\,,
\end{eqnarray}
which is of the same form as Eq.~\eqref{dim6A} modulo the higher-order correction. For now, we will not remove the $\mu^2 C_{DD}$ piece and keep the form of \eref{dim6eom}, mainly because it provides a cleaner relation between $\mathcal L^{(\mathrm{EOM})}_{6}$ and the derivative of the scalar potential.

\subsection{Zero-temperature phenomenology}\label{EDM}
We now discuss experimental constraints on the dimension-six Lagrangians. We begin with the Lagrangian in scenario \textbf{A}. We assume the scalar field picks up a vacuum expectation value (vev) $v_0=246$ GeV, and work in this section in the unitarity gauge $\varphi = (0,\, v_0+h)^T/\sqrt{2}$, where $h$ denotes the Higgs boson with zero-temperature mass $m_H^2 \simeq 125$ GeV. Because of the modified scalar potential, in both scenarios the relations between the parameters $\mu^2$ and $\lambda$ on the one hand and $v_0$ and $m_H^2$ on the other, are modified by the $\kappa$ term. At zero temperature we can express
\begin{equation}
\mu^2 = -\frac{1}{2}\left(m_H^2 - \frac{3}{2}\kappa v_0^4\right)\,,\qquad
\lambda = \frac{1}{2}\left(\frac{m_H^2}{v_0^2}-3\kappa v_0^2\right)\,.\label{mulambda}
\end{equation}
Effects of the dimension-six $\kappa$ interaction in particular induce deviations of the Higgs cubic and quartic interactions with respect to SM predictions. This manifests in processes such as double Higgs production, see e.g. Refs.~\cite{DiVita:2017eyz, Murphy:2017omb} for recent discussions. At the moment, such processes have not been accurately measured and current constraints on $\kappa$ are weak. 

 In scenario \textbf{A}, the dimension-six term in Eq.~\eqref{dim6A} gives a contribution to the top mass. We define the real top mass by
\begin{equation}
m_t = \frac{v_0 y_t}{\sqrt 2} \left( 1+ \frac{v_0^2}{2}C_Y  \right)\,. 
\end{equation}
Although this relation implies that $y_t$ obtains a small imaginary part $\sim \mathcal O(\Lambda_{CP}^{-2})$, this imaginary part only enters observables at $\mathcal O(\Lambda_{CP}^{-4})$ which can be neglected. As such, from now on we use $y_t = \sqrt{2} m_t/v_0 \simeq 1$. 
The interactions between top quarks and Higgs bosons become
\begin{eqnarray}\label{LhA}
\mathcal L_h^{(\mathrm{A})} &=& -\frac{m_t}{v_0} \bar t_L t_R\,h - m_t C_Y\, \bar t_L t_R\left( v_0 h + \frac{3}{2}h^2 + \frac{1}{2}\frac{h^3}{v_0}\right)+\mathrm{h.c.} \nonumber\\
&=& -\frac{m_t}{v_0}\bar t t\,h - m_t \tilde c_{Y}\,\bar t\,i\gamma^5t\left( v_0 h + \frac{3}{2}h^2 + \frac{1}{2}\frac{h^3}{v_0}\right)\,.
\end{eqnarray}

\begin{figure}
\begin{center}
\includegraphics[scale=1]{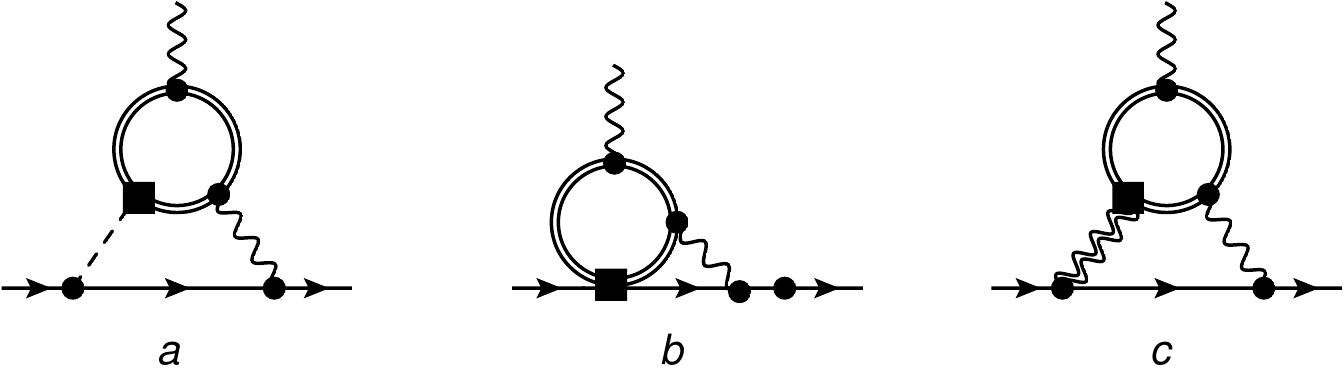}
\end{center}
\caption{Two-loop diagrams contributing to the electron EDM. Single (double) lines denote the electrons (top quarks), dashed lines the Higgs boson, and wavy single (doubles) lines the photons (Z-bosons).  Circles denote SM vertices, while squared denotes CPV dimension-six vertices. Only one topology for each diagram is shown.
}\label{BZ}
\end{figure}

The top-Higgs interactions pick up a CPV component which can be probed in EDM experiments. In particular, the strongest constraint comes from the ACME experiment using the polar molecule ThO, which sets a strong limit on the electron EDM\footnote{This limit assumes negligible contributions to the ThO observable from CPV semi-leptonic operators. This is justified in our scenarios as these semi-leptonic operators are only induced at loop level and strongly suppressed by small Yukawa couplings.} $d_e \leq 8.7 \times 10^{-29}\,e\,\mathrm{cm}$ at $90\%$ c.l. \cite{Baron:2013eja}. The dominant contribution to the electron EDM from the CPV top-Higgs couplings arises from the two-loop Barr-Zee diagram\footnote{We neglect diagrams where the internal photon is replaced by a $Z$-boson. These are suppressed by the electron-Z vector coupling $\sim(-1/4 + \sin^2 \theta_W)$, where $\sin^2 \theta_W \simeq 0.23$ is the square of the sine of the Weinberg angle. }  in Fig.~\ref{BZ}a \cite{Barr:1990vd} and is given by
\begin{equation}\label{deA}
\frac{d^{(\mathrm{A})}_e}{e} = -\frac{32 N_c}{9} \frac{\alpha_{em}}{(4\pi)^3} g(x_t)\,m_e\,\tilde c_Y\,,
\end{equation}
in terms of the number of colors $N_c=3$, the electron mass $m_e$, $x_t = (m_t/m_H)^2$, and the two-loop function
\begin{equation}
g(x_t) = \frac{x_t}{2}\int_0^1 dx\,\frac{1}{x(1-x)-x_t}\log\left(\frac{x(1-x)}{x_t}\right)\simeq 1.4\,.
\end{equation}
The electron EDM limit then sets the strong constraint $|v_0^2 \tilde c_Y| < 0.01$. If we assume $|\tilde c_Y| = \Lambda^{-2}_{CP}$, we obtain $\Lambda_{CP} > 2.5$ TeV. 

In scenario \textbf{B}, the analysis is slightly more complicated. After electroweak symmetry breaking and assuming a purely imaginary $C_{DD}$, the CPV operators  relevant for the EDM calculation become
\begin{eqnarray}\label{CPVB}
\mathcal L_h^{(\mathrm{B})} &=&  - \frac{\alpha  \tilde c_{DD}}{\sqrt{2}}\,\bar t i \gamma^5 t \left( D^2 h  + \frac{m_e}{v_0} \bar e e\right)  - \frac{\alpha  \tilde c_{DD}}{\sqrt{2}}\,\bar t  t \,\bar e i\gamma^5 e\nonumber\\
&=&  - \frac{\alpha  \tilde c_{DD}}{\sqrt{2}}\,\bar t i \gamma^5 t \left( \partial^2 h  + \frac{m_e}{v_0} \bar e e\right) - \frac{\alpha  \tilde c_{DD}}{\sqrt{2}}\bar t t\left(M_Z \partial^\mu Z_\mu + \bar e i\gamma^5 e\right)+\dots\,,
\end{eqnarray}
in terms of the Z-boson mass, $M_Z$, and the dots denote interactions with two or more gauge bosons, which play no role in the EDM calculation. The last two terms in Eq.~\eqref{CPVB} contribute to diagrams \ref{BZ}b and \ref{BZ}c and mutually cancel (this was the reason to include the CPV top-electron coupling in Eq.~\eqref{dim6B}). The first two terms contribute to diagrams \ref{BZ}a and \ref{BZ}b. The contributions can be combined by using $k^2/(k^2-m_H^2) = 1+ m_H^2/(k^2-m_H^2)$ inside the loop, and together become
\begin{equation}\label{deB}
\frac{d^{(\mathrm{B})}_e}{e} = -\frac{32 N_c}{9} \frac{\alpha_{em}}{(4\pi)^3} g(x_t)\,m_e\,\left(-\tilde c_{DD}\frac{\alpha m_H^2}{\sqrt{2}v m_t}\right)\,,
\end{equation}
which is of the same form as Eq.~\eqref{deA}, but with the replacement $\tilde c_Y\rightarrow - (\alpha m_H^2)/(\sqrt{2}v_0 m_t)\tilde c_{DD}$. By specifying $\alpha$, we can ensure the same electron EDM predictions in the two scenarios. In what follows below, we will use
\begin{equation}\label{alpha}
\alpha = -\frac{\sqrt{2}m_t v_0}{m_H^2}\,,\qquad  \tilde c_Y = \tilde c_{DD} = \frac{1}{\Lambda^2_{CP}}\,,
\end{equation}  
with the constraint $\Lambda_{CP} > 2.5$ TeV from the limit on the electron EDM.

In Sect.~\ref{EFT} we argued that scenario $\textbf{A}$ and $\textbf{B}$ are the same apart from higher-order corrections. So where are these higher-order corrections in the EDM calculation? To answer this question it is useful to look at Eq.~\eqref{dim6eom}, which is the CPV Lagrangian after applying the EOM to scenario $\textbf{B}$. The physical real top mass is now given by
\begin{equation}\label{mtEOM}
m_t = \frac{v_0 }{\sqrt 2} \left[ y_t  - \alpha C_{DD} \left( \mu^2  + \lambda v_0^2 + \frac{3}{4} \kappa v_0^4\right)\right]=\frac{v_0 }{\sqrt 2}\, y_t \,,
\end{equation}
where the last equality follows from \eref{mulambda}. After setting $\alpha$ to its value in \eref{alpha}, the interactions between top quarks and Higgs bosons become
\be
\mathcal L_h^{(\mathrm{EOM})} = -\frac{m_t}{v_0}\bar t t\,h - m_t \tilde c_{DD}\,\bar t\,i\gamma^5t\left[ v_0 h + \frac{3}{2}h^2 \left(1+\frac{2\kappa v_0^4}{m_H^2}\right) +\frac{1}{2}\frac{h^3}{v_0}\left(1+\frac{12 \kappa v_0^4}{m_H^2}\right)\right] + \cdots\,,
\label{LhEOM}
\ee
where the dots denote terms with four and five Higgs bosons. Comparing this to \eref{LhA}, we see that the dimension-eight corrections, $\sim \tilde c_{DD} \kappa$, only affect interactions with two or more Higgs bosons. These terms only contribute to the electron EDM  at three loops and these contributions are therefore strongly suppressed. As such, as far the EDM phenomenology is concerned, scenarios $\textbf{A}$ and $\textbf{B}$ are essentially identical.

The CPV top-Higgs interactions give rise to the EDMs and chromo-EDMs of light quarks via very similar Barr-Zee diagrams. Another two-loop diagram involving a Higgs exchange inside a closed top-loop connected to external gluons, gives rise to a CPV three-gluon operator, the so-called Weinberg operator \cite{Weinberg:1989dx}.
The quark (chromo-)EDMs and Weinberg operator in turn give rise to EDMs of the neutron and diamagnetic atoms such as ${}^{199}$Hg and ${}^{225}$Ra. With current experimental sensitivities, these limits are not competitive with the limit from the electron EDM. Furthermore, the hadronic and nuclear EDMs are sensitive to theoretical uncertainties due to hadronic and nuclear matrix elements. A much more detailed discussion can be found in Refs.~\cite{Chien:2015xha,Cirigliano:2016nyn}. 

Finally, the CPV top-Higgs coupling can be directly probed in collider experiments, see e.g. Refs. \cite{Dolan:2014upa,Kobakhidze:2016mfx,Coleppa:2017rgb, Huang:2015izx}. However, for the foreseeable future, the resulting limits are significantly weaker than EDM constraints \cite{Cirigliano:2016nyn}.


\section{The electroweak phase transition}\label{bubble}

\subsection{The finite-temperature Higgs potential}
For the measured value of the Higgs mass, the EWPT in the SM is a cross-over such that the Sakharov condition demanding an out-of-equilibrium process is not satisfied \cite{Kajantie:1995kf,Rummukainen:1998as,Csikor:1998eu}. We have supplemented the Higgs potential in both scenarios therefore by an effective dimension-six operator. In this section we work in the Landau gauge and define the components of the Higgs field as
\be
	\varphi =  
\frac1{\sqrt{2}}
\left( \begin{array}{c}
\theta_1 + i \theta_2  \\ \phi_0+h + i \theta_3 
\end{array} \right)\,,
\ee
with $\theta_i$ the Goldstone bosons, $h$ the Higgs field, and
$\phi_0$ the background field, the tree-level classical potential in terms of $\phi_0$ is given by
\be
	V_0 = \frac{\mu^2}{2} \phi_0^2 + \frac{\lambda}{4} \phi_0^4 +\frac{\kappa}{8} \phi_0^6\,.\label{V0}
\ee 
In order to describe the phase transition we need to include loop corrections to the potential.  The one-loop effective potential can be split into the zero-temperature Coleman-Weinberg potential and the finite-temperature contribution. The former can be resummed to get the renormalization group improved effective potential where the couplings are running with scale.  For the analysis of EWBG we use the coupling values at the renormalization scale $M=m_Z$, and for simplicitly neglect all running effects and threshold corrections. The calculation of the  finite temperature contribution $V_T$ is reviewed in Appendix~\ref{A:thermal}.    We can then write the one-loop effective potential as $V_{\rm eff}= V_{\rm RG} +V_T$, with $ V_{\rm RG}$ the renormalization-group (RG) improved potential, and
\be\label{VT}
	V_{\rm eff} = \frac{\mu^2}{2} \phi_0^2 + \frac{\lambda}{4} \phi_0^4 +\frac{\kappa}{8} \phi_0^6 + \sum_X n_X \frac{T^4}{2\pi^2} J_B (m_X^2/T^2) -\sum_f n_f \frac{T^4}{2\pi^2} J_F(m_f^2/T^2)\,.
\ee
The sums are over all bosons respectively fermions that couple to the Higgs. We only include the fermion contribution from the top quark. $n_X$ and $n_f$ denote the degrees of freedom and are given by $n_{\{h,\theta,W,Z,t\}} = \{1,3,6,3, 4N_c\}$, with $N_c$ the number of colors. The functions $J_{B,F}$ are given by
 \be
	J_{B,F}\left(m^2/T^2 \right) =\int_0^\infty dk k^2 \log{\[1 \mp e^{\[-\sqrt{k^2 + m^2/T^2} \]} \]}\,,
\ee
with the upper (lower) sign for bosons (fermions).  In the high-temperature expansion (see \eref{Jexpand} for the expansion of $J_B$ and $J_F$) the potential becomes
\be
	V_{\rm eff}  = \left(\frac{\mu^2}{2} + \frac{a_T}{2} T^2 \right) \phi_0^2 + \left( \frac{\lambda}{4} + \frac{b_T}{4}T^2 \right)\phi_0^4 + \frac{\kappa}{8}\phi_0^6 +\O(T)\,,
\label{VTexpand}
\ee
where
\begin{equation}
a_T = \frac{1}{16}\left(\frac{4 m_H^2}{v_0^2} + 3g^2 + g'^2 + 4 y_t^2 -12v_0^2 \kappa \right)\,,\qquad b_T = \kappa\,,
\end{equation}
 with $m_H$ and $v_0$ the zero-temperature Higgs mass and Higgs vev, respectively. 
 For simplicity, we will use this high-temperature expansion to determine the allowed values of $\kappa$, and to find the Higgs profile accros the bubble wall that is used for the calculation of the baryon asymmetry. In addition, we neglect higher-loop corrections due to ring diagrams (usually called daisy resummation), and evaluate all running couplings at the scale of the Z-boson mass, and as mentioned above neglect further running effects and threshold effects.  The results are not significantly different from those obtained with the full potential \cite{Delaunay:2007wb}, in which all these effects are included.  Keeping in mind the main goal of this work -- to compare EWBG in the two scenarios and to study the validity of the SM-EFT framework -- here we leave out these complications. For consistency, we compute the thermal corrections to the CPV operator using the same approximations, as discussed in the next section.

 At very high temperatures the effective potential only has a minimum at $\phi_0=0$, while for lower temperatures a second minimum appears. In a potential that allows for a first-order EWPT the two minima are degenerate at some critical temperature $T_c$.  The value of the field $\phi_0$ in the second minimum is denoted by $v_c$.  We find degenerate minima for $\kappa$ in the range $1.6 < (\kappa\times \text{TeV}^{2}) < 4.3$, in agreement with Refs.~\cite{Grojean:2004xa,Delaunay:2007wb}.

The EWPT proceeds by the formation of bubbles of broken vacuum. If larger than some critical size, these bubbles expand and eventually fill up the entire universe. While bubbles can already form at the critical temperature, their rate may be too small for the phase transtion to complete. The temperature at which tunneling to the true vacuum proceeds is called the nucleation temperature $T_N$. To obtain this temperature we follow the discussion in Refs.~\cite{Delaunay:2007wb, Quiros:1999jp}.

The tunneling rate  is $\Gamma \propto e^{-S_E}$, with
$S_E$ the Euclidean action for the so-called bounce solution $\phi_b$ \cite{PhysRevD.15.2929}. At temperatures $T$ greater than the inverse bubble radius $R^{-1}$, the bounce solution is $O(3)$-symmetric \cite{Linde:1981zj} and obeys the equation
\be
  \frac{d^2 \phi_b}{dr^2} + \frac{2}{r} \frac{d\phi_b}{dr} - \frac{\partial V_{\rm eff}(\phi_b,T)}{\partial \phi_b} =0\,,\label{Bounce}
\ee
with boundary conditions
\be
	\phi_b(r\rightarrow \infty)=0 \quad \text{and} \quad \frac{d\phi_b(r=0)}{dr}=0\,.\label{BounceBoundary}
\ee
$\phi_b(r)$ gives the Higgs field profile of a static bubble, with $r$ the distance from the center of the bubble. The corresponding Euclidean action factorizes into $S_E = S_3/T$, with
\be
  S_3 = 4\pi \int dr r^2 \left[\frac{1}{2} \left(\frac{d \phi_b}{dr} \right)^2 + V_{\rm eff}(\phi_b,T) \right]\,.
\ee
Nucleation happens when the probability of creating a single bubble within one horizon is of order one \cite{MORENO1998489}, which leads to the condition
\be
	\frac{S_3}{T_N} \simeq 140\,.
\ee
The value of the field in the true minimum at $T_N$ is denoted by $v_N$.

\begin{figure}
\begin{center}
\includegraphics[scale=0.7]{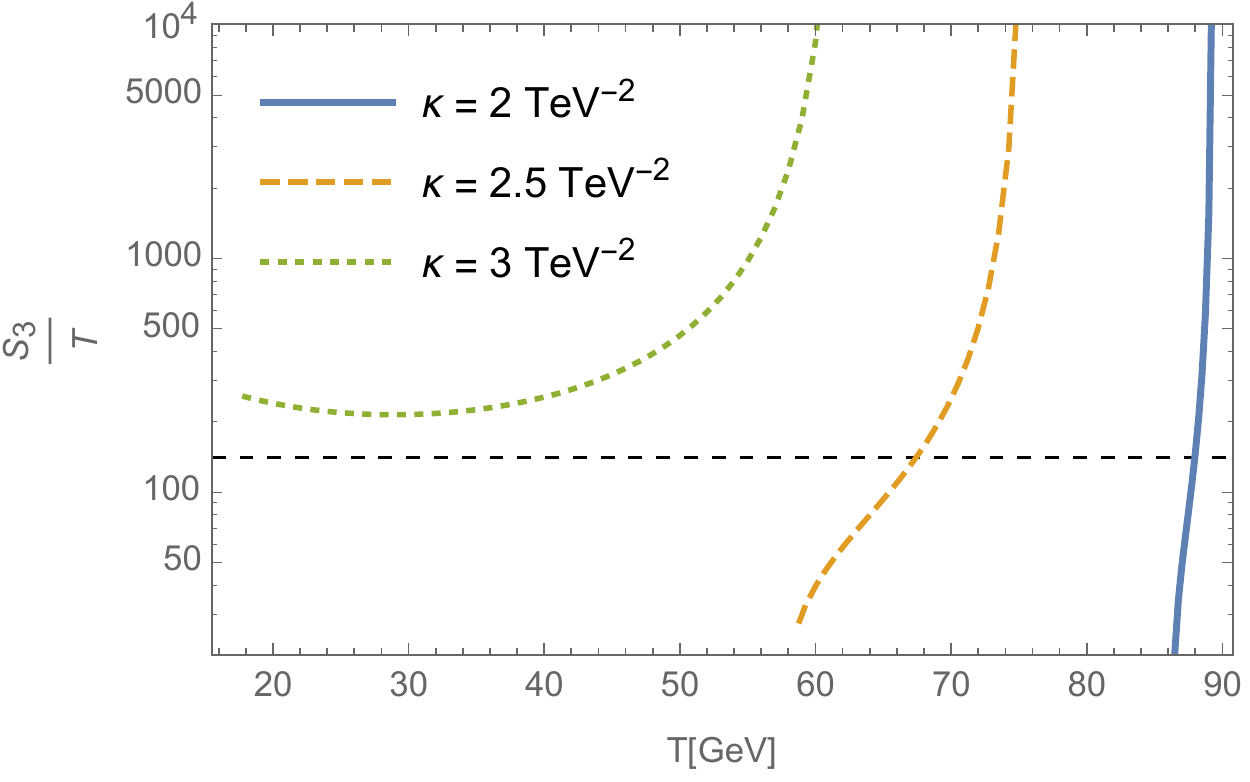}
\end{center}
\caption{${S_3}/{T}$ as a function of temperature for three values of $\kappa$. The horizontal line indicates ${S_3}/{T} = 140$, the approximate value for which bubbles nucleate. The graph shows that nucleation is impossible for $\kappa \gtrsim 3\,\text{TeV}^{-2}$.}\label{NuclTemp}
\end{figure}

We use the Mathematica Package ``AnyBubble" \cite{Masoumi:2016wot} to solve the bounce equation (\ref{Bounce}) and compute $S_3$ for $\kappa = 2, \,  2.5,$ and $3\,\text{TeV}^{-2}$. Fig.~\ref{NuclTemp} shows $S_3/T$ as a function of temperature. For $\kappa \gtrsim 3\,\text{TeV}^{-2}$, the minimum of the potential at $\phi_0=0$ persists until $T=0$, which is reflected in the figure by the lower bound on $S_3/T$. The nucleation rate is never large enough, and $\phi_0$ gets trapped in the symmetric vacuum. For $\kappa \lesssim 1.8\, \text{TeV}^{-2}$ the minimum at $\phi_0 =0$ changes into a maximum before bubbles have had time to nucleate, and the EWPT is not first order. 

In the standard picture of EWBG, a chiral asymmetry is created in front of the bubble wall, which is converted into a baryon asymmetry by sphaleron transitions \cite{Kuzmin:1985mm, Manton:1983nd, Klinkhamer:1984di}. In order to preserve the generated baryon asymmetry in the broken phase, the sphaleron transitions should be suppressed inside the bubble. The rate of sphaleron transitions inside the bubble is proportional to $\exp{[-E_\text{sph}(T_N)/T_N]}$, with sphaleron energy $E_\text{sph}(T_N)$ being proportional to $v_N$.  We therefore demand the additional condition for baryogenesis $v_N/T_N \gtrsim 1$ and refer to Refs.~\cite{Patel:2011th,Fuyuto:2014yia} for a more detailed discussion. We find that this is automatically assured for all values of $\kappa$ for which a first-order phase transition is possible in the first place.  The strength of the phase transition and the value of $v_N/T_N$ increases with $\kappa$.

To summarize, only for a narrow range of values for $\kappa$ do we satisfy all criteria for successful baryogenesis:
\begin{equation}\label{rangekappa}
 1.8 \lesssim (\kappa\times \text{TeV}^{2}) \lesssim 3\,.
\end{equation}  
If we write $\kappa = \Lambda^{-2}$ this corresponds to the scale $0.58 \, \text{TeV}\, < \Lambda < 0.75 \, \text{TeV}\, $.

Finally, we briefly discuss the bubble profile which is needed to calculate the baryon asymmetry.   The bounce solution $\phi_b(r)$ is the initial time ($t=0$) bubble profile. In the rest frame of the bubble, the solution at later times is $\phi_b(\tilde z)$ with $\tilde z = |r - v_w t|$, with $v_w$ the radial velocity of the bubble wall.  We can define a new variable 
\be
z  = r_c - \tilde z = r_c -|r - v_w t| \, ,
\label{zvar}
\ee
with $r_c$ the location of the bubble wall defined via  $\phi_b(r_c) = \phi_b(0)/2$.  In terms of this new coordinate the bubble wall is located at $z=0$, with the broken phase at $z>0$ and the symmetric phase at $z<0$, which matches a convention often used in the literature. We can now write the profile solution $\phi_b(\tilde z(z))$ as a function of $z$.  To calculate the baryon asymmetry  the wall curvature is usually neglected, and the bubble is approximated by a plane located at $z=0$; in this approximation $r$ can be replaced by the coordinate perpendicular to the wall, and $z$ is extended to $\pm\infty$. The value of the bounce solution for $z \rightarrow \infty$ does not exactly equal $v_N$, but has a somewhat smaller value. The difference between $\phi_b(z\rightarrow \infty)$ and $v_N$ is larger when there is a large difference between the potential in the true and the false vacuum.

In the literature the bubble profile is often parametrized by a kink solution \cite{John:1998ip}
\be
	\phi^{\mathrm{kink}}_b = \frac{\phi_b(z\rightarrow \infty)}{2} \left(1 + \tanh{\frac{z}{L_w}} \right)\,,\label{tanh}
\ee
where $L_w$ is a measure of the width of the bubble wall.  The numerical solution can be fit to this parametrization to extract $L_w$.  The kink solution is easy to use, and for scenario \textbf{A} we obtain a baryon asymmetry that only differs from the numerical bounce solution by roughly $10\%$. In scenario \textbf{B}, however, where the baryon asymmetry depends on the Laplacian of $\phi_b$, the kink solution gives very different results. The reason is that the Laplacian contains a term $\frac{2}{\tilde z} \partial \phi_b / {\partial \tilde z}$, which, when integrated over $\tilde z$, is only convergent because of the boundary conditions in \eref{BounceBoundary}, which guarantee that $\partial \phi_b / \partial \tilde z$ goes to zero at $\tilde z=0$. The kink solution, however, does not satisfy the boundary condition exactly and consequently the integral diverges.  The divergence may be tamed by a suitable regulator\footnote{For example, one can add an extra term to the tanh-profile in \eref{tanh} that is small in the bubble wall region, but cancels the divergency at the center $z=r_c$.}, but we will not follow this approach here. To avoid the divergence in scenario \textbf{B},  we will not apply the kink solution for the bubble profile, but instead use the numerical bounce solution in Sect.~\ref{discussion}.

The numerical results presented in Sect. \ref{discussion} are for the benchmark bubble profile, with parameters
\be
{\rm Benchmark:} \quad \kappa = 2\,\mathrm{TeV}^{-2}, \quad  T_N = 88\, \text{GeV}\,,\quad v_N = 148\, \text{GeV}, \quad v_w=0.05\,.
\label{benchmark}
\ee
The value for $\kappa$ corresponds to a cutoff scale $\Lambda = 0.71\,\text{TeV}$ and we have checked that other values of $\kappa$ consistent with a first-order EWPT lead to similar conclusions. The value of the numerical bounce solution for $z\rightarrow \infty$ is given by $\phi_b(z\rightarrow \infty) = 144$ GeV. Fitting to the kink solution, we estimate the width of the bubble wall to be $L_w T_N\simeq 9\,$.  In vacuum the bubble wall would expand at the speed of light, but plasma interactions will reduce the bubble wall velocity.  The calculation of $v_w$ is beyond the scope of this paper, we will use the benchmark value given above \cite{Kozaczuk:2015owa,Bodeker:2009qy,Bodeker:2017cim}.


\section{The matter-antimatter asymmetry}\label{asymmetry} 

All three Sakharov conditions needed for the creation of a matter-antimatter asymmetry are present in the two scenarios outlined in Sect.~\ref{EFT}. The first-order EWPT proceeds via the nucleation of bubbles of the new vacuum, which is an out-of-equilibrium process. The left- and right-handed top quarks in the plasma scatter off the bubble wall differently due to the CPV interactions in Eqs.~\eqref{dim6A} and \eqref{dim6B}.  As a result, a chiral asymmetry is built in front of the bubble wall.  The SM sphaleron transitions only act on the left-handed particles, and transform the chiral asymmetry into a baryon asymmetry. The net baryon charge thus created is swept up by the expanding bubble, and remains conserved provided the phase transition is strong enough such that sphaleron transitions are suppressed in the broken phase inside the bubble.

\subsection{Source term}\label{sec:source}

The number densities of the plasma particles in the presence of an expanding bubble are governed by transport equations. The equations for the top quark will include a CPV source term that drives the chiral asymmetry.  Here we will just sketch the derivation, focusing on how this source term depends on the bubble wall profile.  More details can be found in Ref.~\cite{Lee:2004we}, whose methods we follow.

The quantum transport equations are derived in the finite temperature Closed-Time-Path formalism \cite{Schwinger:1960qe,PhysRev.126.329,Bakshi:1963-1,Bakshi:1963-2,Keldysh:1964ud,CHOU19851}. Starting from the Schwinger-Dyson equation a transport equation for the number current of top quarks can be derived
\begin{align}
\partial_\mu j_i^\mu(x) &=  
-\int d^3 z\int_{-\infty}^{x_0} dz_0\
{\rm Tr}\Bigl[ \Sigma_i^>(x,z) S_i^<(z,x)-S_i^>(x,z)\Sigma_i^<(z,x) 
\nn  \\
&\hspace{4cm} +S_i^<(x,z) \Sigma_i^>(z,x) - \Sigma_i^<(x,z) S_i^>(z,)\Bigr]\,,
\label{transport}
\end{align}
with $i=L,R$ for the left- and right-handed top quark respectively.  Here $S^\lambda$ are the fermionic Wightman functions (see \cite{Lee:2004we} for the explicit definitions), and $\Sigma^\lambda$ the corresponding self-energies defined below in Eq.~\eqref{sigma}.

It is easiest to work in the rest frame of the bubble, where the Higgs profile is only a function of $z =r_c-|v_w t -r|$ as given in \eref{zvar}, and we can express all space-time derivatives in terms of $z$-derivatives.  In the diffusion approximation the current can be written as $j^\mu_i = (n_i, - D_{i} \vec \nabla n_i)$ with $n_i$ the number densities and $D_i$ the diffusion coefficient (see Eq.~\eqref{diffusion}). In addition, we neglect the curvature of the bubble wall, and model the bubble wall as a plane located at $z=0$. With these approximations
\be
\partial_\mu j_i^\mu(x) 
\simeq v_w n_i' -D_i \vec \nabla^2 n_i
\simeq  v_w n_i' -D_i n_i''\,,
\label{dJ}
\ee
where the last expression is valid for the planar approximation, and where a prime denotes a derivative with respect to $z$.

In the bubble background the top quark mass is space-time dependent as it depends on the Higgs background $\phi_b(z)$.  To deal with this complication, the self-energies are calculated in the ``vev-insertion approximation'' \cite{Carena:2002ss,Konstandin:2005cd,Cirigliano:2009yt,Cirigliano:2011di}, which amounts to treating the field dependent part of the top mass as a perturbation. To compare the asymmeties produced in scenarios \textbf{A} and \textbf{B} in a consistent way it is important to work at the same order in perturbation theory in both the Higgs and the CPV sector.  Thus we include the one-loop thermal corrections to the CPV interactions, which are calculated in Appendix~\ref{A:thermal}, and neglect daisy diagrams.  The zero-temperature top mass\footnote{In Sect.~\ref{EDM}, we used the symbol $m_t$ to denote the real top mass at zero temperature, which is relevant for the EDM calculation. In the current section, however, $m_t$ is a complex number.} $m_t$ can be split into a real and imaginary part (indicated by superscripts), and likewise for the thermal corrections $\delta m_t$.  The quadratic Lagrangian for the top quarks is split into a free part, independent of the bubble profile, and a field-dependent interaction part, according to
\begin{align}
\mathcal L^{\rm free}
& \supset 
\sum_{i=L,R} \bar t_i \( i\slashchar{D}-\delta m_i^{\rm Re} (T) \) t_i\,,
\label{Lmass} \\  
\mathcal L^{\rm int}
& \supset   - \[ m_t^{\rm Re}(\phi_b) + i \(m_t^{\rm Im}(\phi_b)  +
     \delta m_t^{\rm Im}(\phi_b,T) \) \] \bar t_L t_R+ {\rm h.c.} 
\equiv - \frac{f(T,\phi_b)}{\sqrt{2}} \bar t_L t_R + {\rm h.c.}
\label{Lint}
\end{align}
The $f$-functions defined above, which parameterize the interaction strength, are derived in Appendix \ref{A:thermal} for the scenarios under investigation.
$\delta m_{i}^{\rm Re}(T) $ are the usual SM thermal masses \cite{Enqvist:1997ff}, which we list in \eref{deltamR}.  They can be viewed as one-loop thermal corrections to the massless propagator. Since these corrections do not depend on the space-time dependent Higgs profile, they can be resummed and included in the full propagator $S_i^\lambda(\delta m_{i}^{\rm Re})$, which is constructed from the free Lagrangian.  $\delta m_{i}^{\rm Im}(\phi_b,T)$ are the one-loop thermal corrections to the CPV $m_{i}^{\rm Im}$-vertex.  All the terms in $f$ are field dependent, and therefore treated as a perturbation. The imaginary part of the top mass is space-time dependent in the bubble background, and cannot be rotated away by a chiral transformation if it is non-linear in the field. Its presence leads to different dispersion relations for left- and right-handed particles, and consequently different forces act on them as they scatter with the bubble wall.  This is the physical underpinning of the appearance of a source term, denoted by $S^{\CPV}$, in the transport equations that drives the chiral asymmetry.  Based on this discussion, we expect $S^{\CPV} \propto {\rm Im}(f' f^*)$, as it should be proportional to $f'$, depend on the phase of $f$, and be quadratic in $f$ as the diagram for $t_L \to t_L$ scattering requires at least two mass insertions. This is confirmed by the explicit derivation, which we will now sketch.

We consider the transport equation for the right-handed top quark $t_R$. The self-energy $\Sigma^\lambda_R$ obtains a contribution from the diagram with two mass insertions
\be
\Sigma_R^\lambda(x,y) = - f(x) f^*(y) P_R S_L^\lambda(x-y) P_L\,,
\label{sigma}
\ee
with $P_{L,R}$ the left- and right-handed projection operators.  Using \eref{sigma} in the transport equation, \eref{transport}, we can separate the right-hand side into a real and imaginary part, corresponding to the CP-conserving relaxation term and the CPV source
\begin{align}
\partial_\mu j_R^\mu(x) 
&= \frac12 \int d^3z\int_{-\infty}^{x_0} dz_0\
\left[f_xf_z^*+f_x^*f_z\right] \ {\rm Re} \,
{\rm Tr}\left[S_L^>(x,z) S_R^<(z,x)
-S_L^<(x,z) S_R^>(z,x)\right]_{{\rm Tr} (m)=0} 
\nn \\
&+
\frac12 \int d^3z\int_{-\infty}^{x_0} dz_0\,
 i\left[f_xf_z^*-f_x^*f_z\right]\ {\rm Im}  \,
{\rm Tr} \left[S_L^>(x,z) S_R^<(z,x)
-S_L^<(x,z) S_R^>(z,x)\right]_{{\rm Tr} (m)=0} 
\nn \\
&= S_R^{{\rm CP}}(x) +  S_R^{\CPV}  (x)\,,
\label{split}
\end{align}
where we used the short-hand $f_x = f(x)$. The subscript ${{\rm Tr} (m)=0} $ indicates that mass can be set to zero in the trace of the propagators\footnote{Inserting \eref{sigma} in \eref{transport} gives a trace of a product of propagators and projection operators, which in Fourier space is of the form
  ${\rm
    Tr}\[P_L (\slashed{p} + m_i) P_R (\slashed{q} + m_j)\]=\frac12{\rm
    Tr}\[\slashed{p}\slashed{q} \]$.  By defining   
\be
S_i^\lambda(x) \big|_{{\rm Tr} (m)=0} = \int \frac{\dd^4 k}{(2\pi)^4} \e^{i k.x} \hat S_i^\lambda(k) (\slashed{k}+m) \bigg|_{{\rm Tr} (m)=0} 
=  \int \frac{\dd^4 k}{(2\pi)^4} \e^{i k.x} \hat S_i^\lambda(k) \slashed{k}\,,
\ee
\eref{split} can be neatly split into a CP-conserving and CPV part.}.
The analagous equation can be written down for the left-handed quark,
with $S_L^{{\rm CP}}(x)=- S_R^{{\rm CP}}(x) $ and
$S_L^{\CPV}= -S_R^{\CPV} $.

In the limit that the typical time scale for thermalization of the top
quarks is much faster than the time scale on which the Higgs profile
changes, we can expand\footnote{Here we used that Taylor expanding $\lim_{z\to x} \left[f_xf_z^*-f_x^*f_z\right]$,  the $f^* \partial_i f$ term vanishes when substituted in the integral in \eref{split} because of spatial isotropy, and thus only the term proportional to the time-derivative  $f^* \partial_0 f =v_w f^*  f'$ contributes \cite{Lee:2004we}.} 
\be 
\lim_{z\to x} \left[f_xf_z^*+f_x^*f_z\right]  \approx 2|f(x)|^2\,,
\qquad
\lim_{z\to x} \left[f_xf_z^*-f_x^*f_z\right]  \approx 2 i v_w \Im( f'(x)
f(x)^*) (x^0-y^0)\,,
\ee
and the $f$-dependent parts can be taken outside the $z$-integral in \eref{split}. This gives the result we are after, as it factors out the explicit dependence on the bubble-wall profile.  We thus find that $S_R^{{\rm CP}} \propto |f|^2 $ and $S_R^{\CPV} \propto \Im( f' f^*) $, with the constant of proportionality a function of the temperature, thermal masses $\delta m_i^{\rm Re}$, and top decay width only, as these are the quantities entering the propagator.  Moreover, the thermal corrections to the CPV operator, and thus to the source, and the effective potential are calculated consistently. 


\subsection{Transport equations}

To calculate the chiral asymmetry in front of the bubble wall we keep track of the number density of the third-generation quarks and the Higgs field.  The electroweak gauge interactions are fast, and approximate chemical equilibrium between the members of the left-handed doublet is assumed.  Consider then the following densities $Q= n_{t_L} + n_{b_L}$, $R = n_{t_R}$, and $H= n_h$, with $n_i$ the number density of quarks minus anti-quarks, and for the real Higgs field the number density of Higgs particles. Since the CP violation resides purely in the top quark sector\footnote{We neglect the CPV top-electron coupling that appears in scenario \textbf{B} (see \eref{dim6B}) as it is proportional to the small electron Yukawa coupling.}, no asymmetry is built up in the lepton sector. The first- and second-generation quarks only interact via strong sphaleron processes on the relevant time scales, and their densities can be related to those of the third generation.  The total chiral asymmetry is $n_L = 5Q +4R$ \cite{Huet:1995sh}. Because of the different time scales involved we can describe the creation of the chiral asymmetry, and the transformation into a net baryon asymmetry as a two-step process.

The set of coupled transport equations can be derived as explained in the previous section. In addition to the relaxation and source term from the mass-insertion diagrams, there are Yukawa interactions that contribute to $S^{\rm CP}$.  The (non-perturbative) strong sphaleron interactions are also included. The full set of transport equations is \cite{Lee:2004we}
\begin{align}
\partial^\mu( R+Q)_\mu&= 
- \Gamma_{ss}\left({2Q\over k_Q} -{R\over k_{R}}+{(Q+R)\over
    k_{\rm eff}}\right)\,,
 \nn \\
\partial^\mu H_\mu &= \Gamma_Y\left({R\over
                     k_{R}}-{H\over k_H}-{Q\over k_Q}\right)\, ,
\nn \\
\partial^\mu (2R+Q+H)_\mu &=  \Gamma_M^{+} \left({R\over k_{R}}+
{Q\over k_Q}\right)-\Gamma_M^{-} \left({R\over k_R}-{Q\over
  k_Q}\right)
+ S_R^{\CPV} \,.
\label{cascade}
\end{align}

All rates and input parameters needed to solve this set of equations are given in Appendix~\ref{A:input}.  $\Gamma_{ss}, \, \Gamma_Y$, and $\Gamma_M^\pm$ are the strong sphaleron rate, the Yukawa interaction rate, and the relaxation rate, respectively.  The latter two are extracted from $S^{\rm CP}$ and, as discussed in the previous subsection, are proportional to $|f|^2 = (y_t \phi_b)^2 + \O(\Lambda_{CP}^{-4})$. The difference between scenario \textbf{A} and \textbf{B} lies thus solely in the source term, which we give here explicitly\footnote{In
  the expressions for $\Gamma_Y , \, \Gamma_M^\pm$,  and $S^{\CPV}$ we have
  neglected the collective plasma hole excitations to the
  propagators \cite{Weldon:1989ys,Weldon:1999th,Klimov:1981ka}.}
\begin{align}
S^{\CPV} &= 
\frac{N_c v_w}{\pi^2} \Im\(f' f^*\) 
  \int \frac{k^2 \dd k }{\omega_L
  \omega_R} \Im \bigg[ \nn
\frac{(n_f(\E_L) -n_f(\E_R^*))}{(\E_L -\E_R^*)^2}
\( \E_L \E_R^* -k^2\) \nn \\
& \hspace{5cm}+ \frac{ (n_f(\E_L) +n_f(\E_R)-1) }{(\E_L +\E_R)^2}
\( \E_L
 \E_R +k^2\)\bigg]\,,
\end{align}
where $n_f(x) = (\e^x+1)^{-1}$ denotes the Fermi-Dirac distribution, $\E_i = \sqrt{k^2 +(\delta m_i^{\rm Re})^2} -i \Gamma_t$, and $\Gamma_t$ the top decay width.  The ``-1'' term in the numerator on the second line gives a divergent contribution that survives in the zero-temperature limit where the distributions $n_f$ are Boltzmann suppressed.  This divergence is absorbed by the counterterms of the zero-temperature renormalized action, or equivalently, this term can be removed by normal ordering the operators \cite{Liu:2011jh}.

Assuming local thermal equilibrium and small chemical potentials, the
$k$-functions are implicitly defined via (see \eref{kfun} for more details)
\be
n_i= \frac{\mu_i T^2}{6} k_i (m_i/T) + \O(\mu_i^3)\,,
\ee
where the mass can be approximated by the real part
$m_i = m_i^{\rm Re} + \delta m_i^{\rm Re}$
\footnote{Since the r.h.s. of the transport equation is calculated using the vev-insertion approximation, it can be argued that the mass used in the $k_i$-functions should be the thermal mass instead, i.e. $k_i(\delta m_i^{\rm Re})$.  Doing so would only give a small difference in the final asymmetry.}.
Furthermore we have
\be
k_{\rm eff} \equiv \(\frac{4}{k_{Q_{1L}}}+
   \frac{4}{k_{Q_{2L}}}+\frac{1}{k_{U_R}}+\frac{1}{k_{C_R}}+\frac{1}{k_{D_R}}+\frac{1}{k_{S_R}}+\frac{1}{k_{B}}\)^{-1}\, ,
\label{keff}
\ee
which is often approximated by $1/k_{\rm eff} \simeq 9/k_B$ \cite{Huet:1995sh}.

The set of transport equations \eref{cascade} reduces to ordinary
differential equations in the approximation of \eref{dJ}, and can be
solved to find the net chiral assymmetry $n_L =5Q+4T$.  The SM
sphalerons convert this into a net baryon number. Integrating over the
asymmetric phase $z <0$, where the sphalerons act, the baryon asymmetry becomes 
\be
\label{nb_sol}
Y_B = \frac{n_b}{s}= - \frac{3 \Gamma_{\rm ws}}{2 v_{w}s} \, \int_{-\infty}^{0}  \ 
n_L (z) 
\,  e^{z \, {\cal R} \Gamma_{\rm ws}/v_w} \, 
\dd z \ . 
\ee
which is to be compared with the data in \eref{BAUExp}.
Here $s = 2 \pi^2/(45) g_{*S} T^3$ is the entropy density, and $g_{*S} = 106.75$ the entropy degrees of freedom at the electroweak scale. The relaxation term ${\cal R} = 15/4$ in the SM, and the weak sphaleron rate is $\Gamma_{\rm ws} = 6 \kappa \alpha_w^5 T$ with $\kappa \sim 20$ and $\alpha_w = g^2/(4\pi)$.  


\section{The baryon asymmetry and investigation of the SM-EFT expansion}\label{discussion}

In this section we compare the baryon asymmetry computed in scenario {\textbf A} and {\textbf B}, and use this as guidance to investigate the validity of the SM-EFT expansion. We numerically compute the baryon asymmetry, using the methods described in Sect.~\ref{asymmetry}. We use the benchmark Higgs profile described by \eref{benchmark}.

\subsection{Interaction strength and source term}
\label{deltasection}

 An important ingredient in the calculation of the asymmetry is the interaction strength between left- and right-handed top quarks $f(T,\phi_b)$  defined in \eref{Lint}, which depends on the temperature, the bubble profile, and on the source of CP violation. The various interaction strengths $f_i(T,\phi_b)$, where $i=\{A,B,\mathrm{EOM}\}$ corresponding to the CPV operators in scenarios \textbf{A} (\eref{dim6A}), \textbf{B} (\eref{dim6B}), and \textbf{B} after applying the EOM (\eref{dim6eom}), have been calculated in Appendix~\ref{A:f} in the high-temperature limit and are given explicitly in \eref{fnum}.  The baryon asymmetry in particular depends on the combination $S^{\CPV} \propto \delta_i \equiv\mathrm{Im}(f'_i f^*_i)$ which enters the source term, and varies between the scenarios.
 
 In order for $\delta_i$ to be nonzero, we require that $f_i$ has both real and imaginary parts, and $f_i$ must have at least one term that depends non-linearly on the background field. For instance, for a linear dependence, $f = c\,\phi_b$ with $c$ any complex number, it is clear that $\delta_i=\mathrm{Im}(|c|)\phi_b \phi_b' =0 $. The requirement of a non-linear dependence reflects that a CP-phase in the SM dimension-four Yukawa term can be rotated away and is not physical, see also the discussion surrounding  \eref{dim6eom2}.  With these considerations we obtain in the different scenarios
\begin{align}
 \delta_A& =  \tilde c_Y y_t^2 \phi_b^3 \phi_b' \, ,\label{fprimefstarA}\\
 \delta_B &= -\alpha y_t \tilde c_{DD} \(\phi_b \phi_b^{'''} -\phi_b^{'} \phi_b^{''}+ \frac{2}{z-r_c}\phi_b\phi_b^{''}  -\frac{2}{z-r_c}\phi_b^{'2} -\frac{2}{(z-r_c)^2}\phi_b\phi_b^{'}\),\label{fprimefstarB} \\
  \delta_{\mathrm{EOM}} &= -\alpha\,y_t \tilde c_{DD} \[ 2\(\frac{m_H^2 - 3v_0^4\kappa}{2v_0^2} + \kappa T^2\)  \phi_b^3 \phi_b^{'} + 3\kappa  \phi_b^5 \phi_b^{'} \] \,.\label{fprimefstarEOM}
\end{align}
The results for $f_\text{EOM}$ were obtained by first applying the EOM to the tree-level Lagrangian of scenario {\textbf B}, and then calculating the one-loop thermal corrections to $f_\text{EOM}$. We have checked in Appendix~\ref{A:thermal} that the same result is obtained if we first  calculate  the thermal corrections in scenario {\textbf B} to determine $f_B$, and then apply the one-loop equations of motion to obtain $f_\text{EOM}$.  Since the bounce solution $\phi_b$ is a solution of the one-loop equations of motion, see \eref{Bounce}, it follows that  $f_\text{EOM} = f_B$ and the source terms are equivalent --- just as expected. This is somewhat obscured by the form of $\delta_i$ given above, but it is apparent from the $f$-functions given in \eref{ffunctions}.  For this comparison, we stress that it is important to  consistently include thermal corrections to the effective potential and to the CPV operators.

The difference between scenario {\textbf A} and {\textbf B} arises from the difference between $\delta_A$ and $\delta_{\mathrm{EOM}}$.  Normalizing the CPV operators such that they give the same EDM constraints by using \eref{alpha}, this gives the relation
\be
\delta_B = \delta_{\rm EOM} = \delta_A \(1+ \kappa \frac{v_0^2}{m_H^2} \( 3(\phi_b^2-v_0^2)+2 T^2\)\) \,.
\label{differ}
\ee
The scenarios thus differ by the terms proportional to $\kappa\, \delta_A$ that scale as $\mathcal O(\Lambda^{-2}\Lambda_{CP}^{-2})$.

\subsection{Baryon asymmetry in scenario \textbf{A} and \textbf{B}}
\label{asymsection}
\begin{figure}
\begin{center}
\includegraphics[scale=.6]{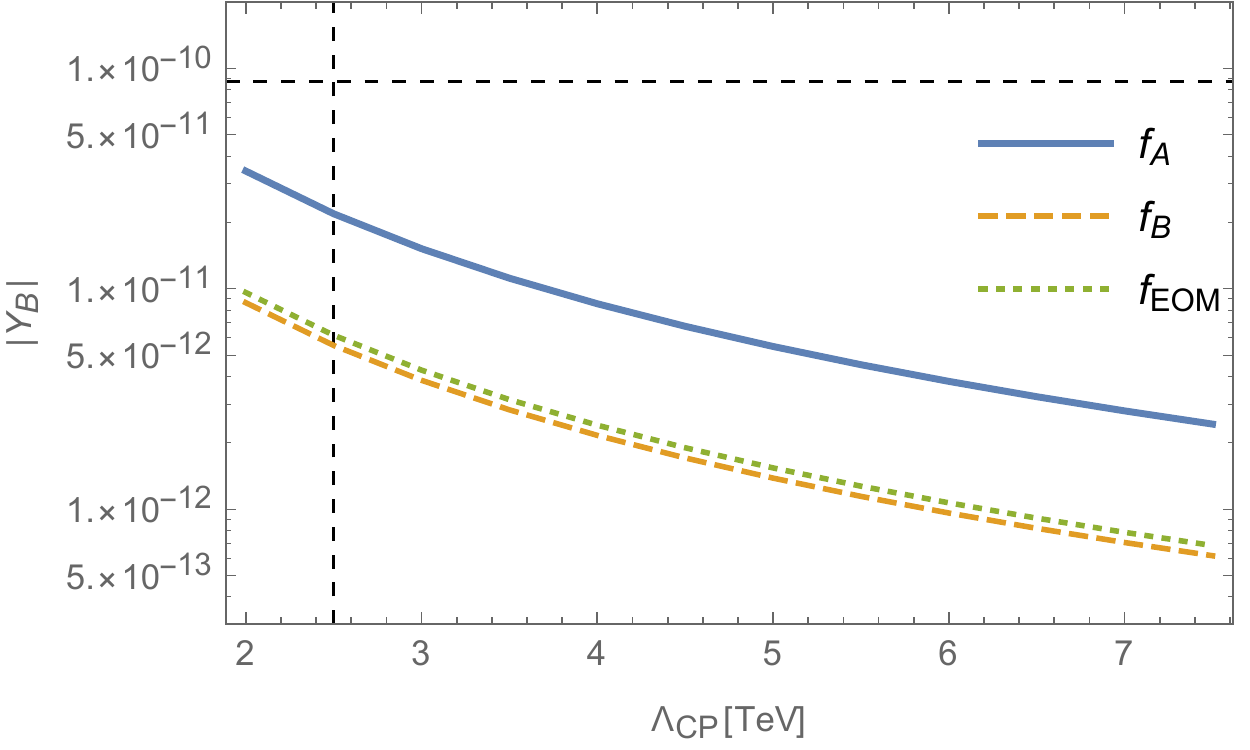}
\includegraphics[scale=.6]{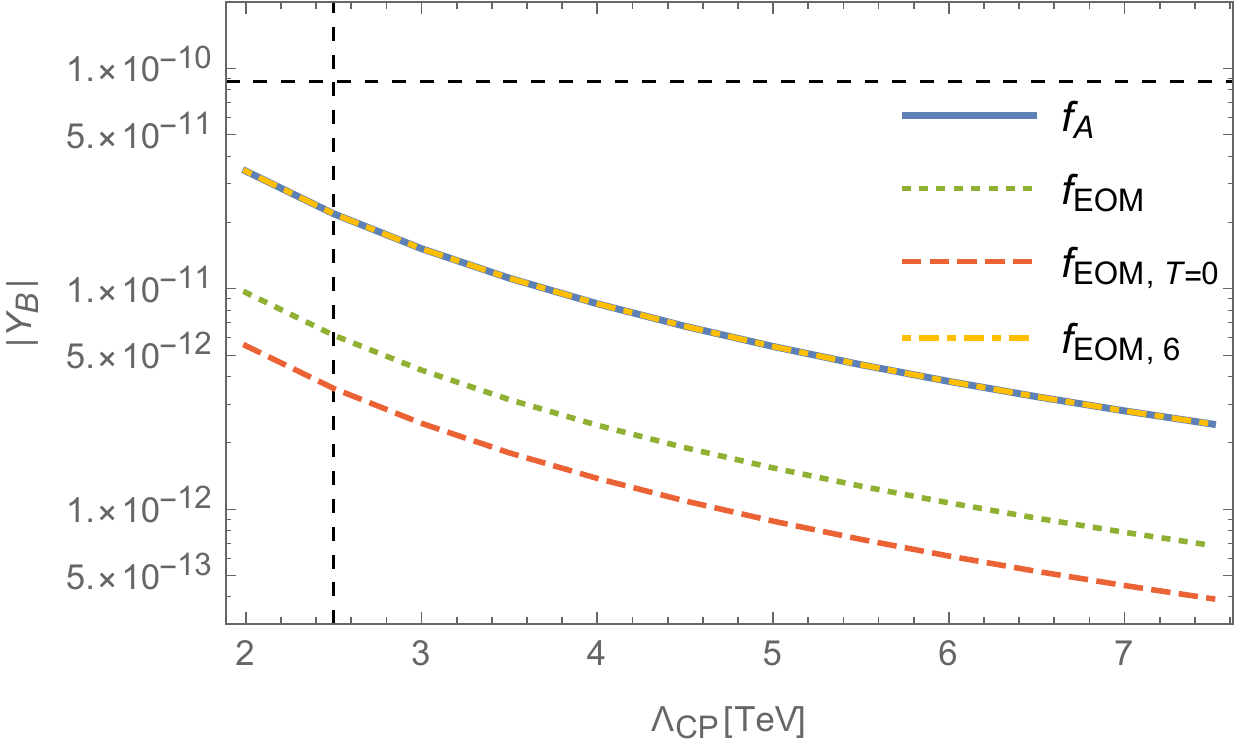}
\end{center}
\caption{Absolute value of the baryon asymmetry for the considered dimension-six operators as a function of the effective CPV scale $\Lambda_{CP}$. The vertical line indicates the experimental cutoff on $\Lambda_{CP}$ and the horizontal line indicates the observed value of the baryon asymmetry.}\label{BAUcutoff}
\end{figure}

We will now discuss the baryon asymmetry in the two scenarios, starting with scenario \textbf{A}. We solve the transport equations in \eref{cascade} with the semi-analytic method outlined in Appendix~\ref{A:solution} \cite{White:2015bva}, and feed the solution for $n_L(x)$ into \eref{nb_sol} to obtain the baryon asymmetry.  We plot the baryon asymmetry as a function of $\Lambda_{CP}$ in Fig.~\ref{BAUcutoff} in solid blue. The asymmetry measured by PLANCK is depicted by the dashed horizontal line, while the constraint on $\Lambda_{CP}$ from EDM experiments ($\Lambda_{CP}>2.5$ TeV) is depicted by the vertical dashed line. The asymmetry is proportional to the amount of CP violation, $Y_B \propto \tilde c_Y =\Lambda_{ CP}^{-2}$, thus raising the cutoff scale by a factor $\sqrt{2}$ will decrease the asymmetry by a factor two.   Our results indicate that, given our approximation and input values, scenario \textbf{A} cannot produce the observed asymmetry for cutoff scales consistent with the EDM experiments.  For the lowest allowed scale $\Lambda_{CP} =2.5$ TeV, the asymmetry is too small by roughly a factor $4$.

Despite this too small asymmetry we make no definite statements on the viability of scenario \textbf{A} because our result for $Y_b$ should be taken with a grain of salt. The baryon asymmetry is calculated using several approximations. For fermionic CPV sources in particular there are still a number of outstanding problems, see for example Refs.~\cite{Cirigliano:2006wh, Morrissey:2012db}. Other issues are related to the accuracy of the vev-insertion approximation and the high-temperature expansion we applied in the calculation of the effective potential and the CPV source \cite{Curtin:2016urg}, and the uncertainty in the bubble wall velocity \cite{Kozaczuk:2015owa,Bodeker:2009qy,Bodeker:2017cim}.  We therefore stress that our result for the baryon asymmetry suffers from significant theoretical uncertainties. As the main goal of this work is to study the SM-EFT framework in the context of EWBG, we can live with these uncertainties as we are not too interested in the exact value of $\Lambda_{CP}$ necessary for successful baryogenesis. 
An improvement of the formalism such that more accurate predictions can be made would be very relevant, in particular considering that next-generation EDM experiments aim for an order-of-magnitude improvement in sensitivity.

We now turn to scenario \textbf{B}. In this scenario the source is proportional to the much more complicated expression in \eref{fprimefstarB} and depends also on the second and third derivative of the bubble profile.  The baryon asymmetry is obtained in the same way as in scenario \textbf{A} and it is plotted in Fig.~\ref{BAUcutoff} in dashed yellow. The asymmetry is roughly four times smaller than in scenario \textbf{A} for the same values of $\Lambda_{CP}$.  Although $Y_B$ suffers from the same uncertainties described above, it is fair to say that the tension with EDM constraints is significantly larger in scenario \textbf{B}.

Our results are obtained for the specific value of $\kappa$ in \eref{benchmark}, which is expected to be representative for the narrow range in \eref{rangekappa} consistent with a first-order EWPT. We have checked that our results do not change qualitatively by small modifications of $\kappa$, but the exact difference between the value of the baryon asymmetry in scenarios \textbf{A} and \textbf{B} does depend on $\kappa$.

The large difference in the baryon asymmetry in the two scenarios points towards a breakdown of the SM-EFT expansion and indicate that the higher-order corrections in \eref{differ} play an important role even though the power counting suggests that such effects are suppressed by $\mathcal O(v_N^2/\Lambda^2)$. In the next subsection we investigate the higher-order corrections in more detail.

\subsection{Thermal corrections and dimension-eight effects}
We study the difference between scenario \textbf{A} and \textbf{B} by dissecting $\delta_{\text{EOM}}$ in \eref{fprimefstarEOM}, which drives the CPV source after applying the EOM on scenario \textbf{B}. If we consider the complete $\delta_{\text{EOM}}$, consisting of dimension-six and -eight (proportional to $\kappa$) contributions, we obtain the baryon asymmetry that is plotted in the left and right panel of Fig.~\ref{BAUcutoff} in dotted green. As it should, the asymmetry coincides with that of scenario $\textbf{B}$. The small differences of about $10\%$ are due to numerical issues\footnote{The derivative terms in \eref{fprimefstarB} diverge in the limit $z \to r_c$ corresponding to the centre of the bubble, and the result is only finite once all terms are combined. Numerical errors arise if this cancellation is not perfect.} related to the derivatives appearing in $\delta_{B}$. The fact that the asymmetries agree provides a nontrivial check of our calculation.

 If we now turn off dimension-eight contributions in $\delta_{\text{EOM}}$, by setting $\kappa =0$ in \eref{fprimefstarEOM}, we reproduce the asymmetry in scenario \textbf{A}, as illustrated by the dotted-dashed yellow line in the right panel of Fig.~\ref{BAUcutoff}. This is also expected as scenarios \textbf{A} and \textbf{B} are the same up to dimension-eight effects. The conclusion is that the formally higher-order terms in $\delta_{\text{EOM}}$ proportional to $\sim \kappa \tilde c_{DD}$,  reduce the obtained asymmetry by about a factor four and the EFT expansion explicitly fails. The dimension-eight terms can be separated in a temperature-independent and -dependent piece, and the dashed red line in the right panel of Fig.~\ref{BAUcutoff} is the result when we neglect the temperature-dependent piece of the CPV source. The difference with the full result is now roughly a factor six such that neglecting the temperature corrections to the CPV source is a poor approximation. 
 
 So what causes the breakdown of the EFT expansion? While the scale related to the $(\varphi^\dagger \varphi)^3$ term is not very high, $\Lambda =0.71$ TeV, it is still significantly larger than any other scale, such as $v_0$, $v_N$, $T_N$, or particle masses, appearing in the computation. The problem is related to the demand of a strong first-order EWPT. The presence of a second minimum in the scalar potential requires a detailed balance between the dimension-two and -four terms in the SM Lagrangian and the $\kappa (\varphi^\dagger \varphi)^3$-term. This is the case for the zero-temperature potential as well as for the potential at the nucleation temperature $T_N$, but in the latter case the temperature corrections are included in the balancing act. This spoils the hierarchy between the dimension-four and -six terms in the effective potential, and thus after applying the EOMs for the Higgs field it spoils the hierarchy in the CPV sector.

This problem is not manifest in the EDM predictions in scenario \textbf{A} and \textbf{B}. The EDM constraints only depend on the linear top-Higgs coupling, which when expressed in terms of the physical masses, are the same in the two scenarios, and the higher-order terms (suppressed or not) are irrelevant. However, if we were able to accurately measure, for example,  the CPV $h^2\, \bar t\, i \gamma^5\, t$-coupling, the two scenarios would give different predictions as can be seen by comparing \eref{LhA} and \eref{LhEOM}; supposedly higher-order corrections in the EFT counting of the form $2 \kappa v_0^4/m_h^2 \sim \mathcal O(1)$ give order-one corrections because of the balancing act in the Higgs potential. 

\begin{figure}
\begin{center}
\includegraphics[scale=.6]{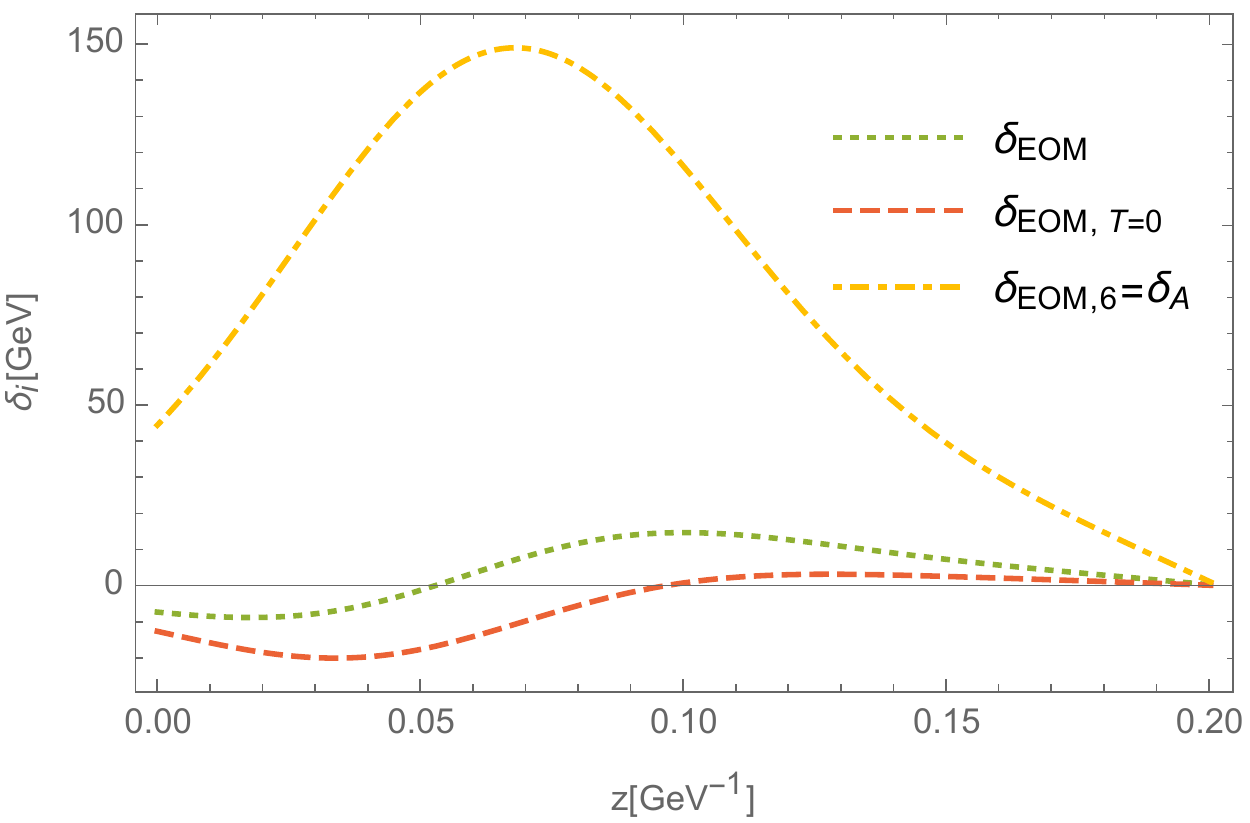}
\includegraphics[scale=.6]{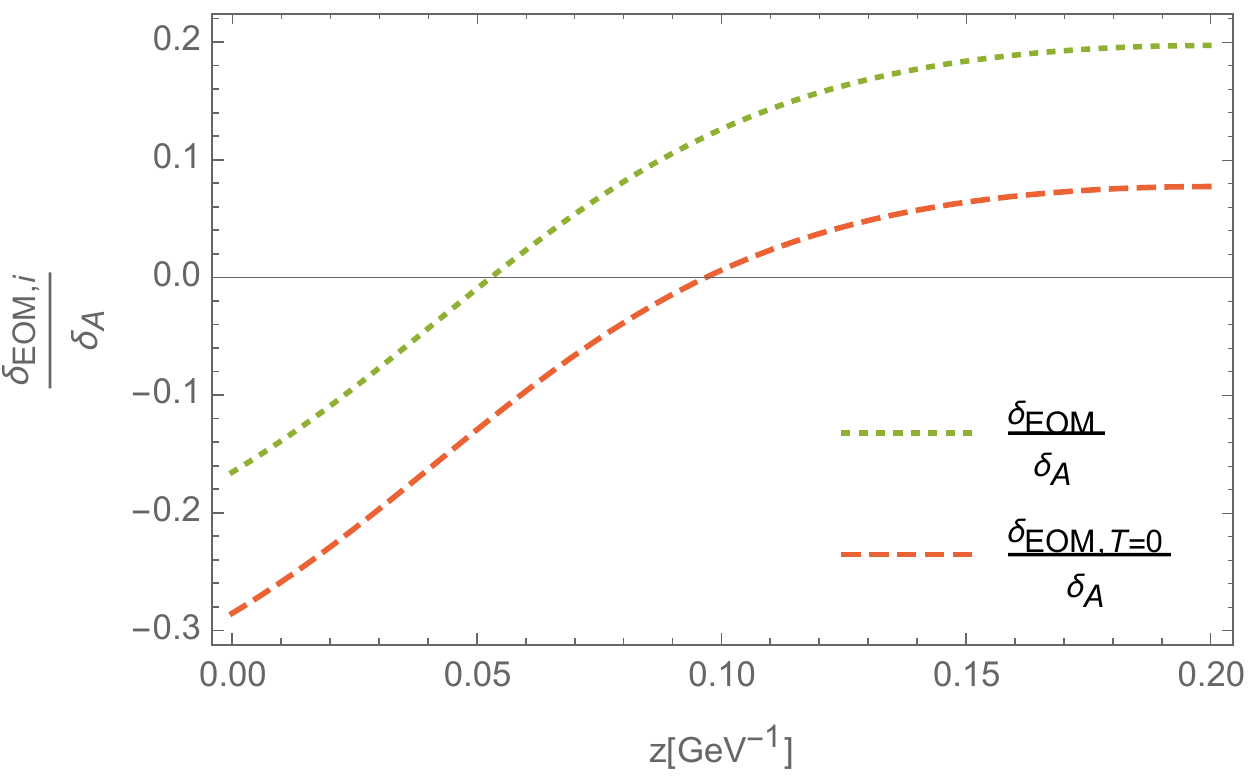}
\end{center}
\caption{Left panel: $\delta_\text{EOM}$ (green, dotted) and $\delta_{\text{EOM},6} = \delta_A$ (yellow, dotted-dashed) as a function of $z$ in the broken phase. The difference arises from the terms proportional to $\kappa T^2\phi_b^3$ and $\kappa \phi_b^5$ in \eref{fprimefstarEOM}.  To see the relative relevance of these two terms, we also plotted $\delta_{\text{EOM},\, T=0}$ (red, dashed), where the thermal corrections are neglected. Right panel: $\delta_\text{EOM}/\delta_A$ in the broken phase. In dotted green the finite temperature corrections are included, in dashed red they are neglected.}
\label{fig:imffstar}
\end{figure}

The same breakdown of the EFT expansion occurs in the calculation of the baryon asymmetry. The difference between the obtained asymmetries in scenarios  \textbf{A} and  \textbf{B} arise from the dimension-eight terms in \eref{differ}. Fig.~\ref{fig:imffstar} depicts the $\delta_{\rm EOM}$ as a function of $z$. The green dotted line is the full result including all terms, the dashed red line ignores the temperature corrections, and the dotted-dashed yellow line ignores all dimension-eight effects and thus coincides with $\delta_A$. The broken phase extends to $z\rightarrow \infty$, but $\phi^{'}_b$ and consequently $\delta_i$ are only nonzero for $z<r_c\simeq 0.2 \,\text{GeV}^{-1}\,$.  The final baryon asymmetry depends on a weighted integral of the source over the broken phase\footnote{See Appendix \ref{A:solution} for more details on the solution to the transport equations. The source enters the solution through the $\beta_i$-parameters as given in Eqs.~(\ref{infinity}) and (\ref{matching}).}.


The difference between the source with and without dimension-eight CPV interactions is not small at all. In fact, the peak value of $\delta_A$ is more than an order of magnitude larger than the peak value of $\delta_{\mathrm{EOM}}$. The right panel of Fig.~\ref{fig:imffstar} shows $\delta_\text{EOM}/\delta_A$ in the broken phase. The ratio goes from roughly $-0.2$ at $z=0$ to $0.2$ at $z=0.2$. The difference mainly arises from zero-temperature contributions to $\delta_{\rm EOM}$, although the temperature corrections are  non-negligible. Apart from the difference in overall scale,  
$\delta_{\mathrm{EOM}}$ has a zero-crossing point which causes a partial cancellation between the positive and negative contributions in the integrand. The zero-crossing point emerges because the various terms in \eref{fprimefstarEOM} are of the same order and have opposite sign, which is related to the necessity of a second minimum in the potential. The cancellation is also present for $\delta_{\mathrm{EOM},\,T=0}\,$, but occurs at a different value of $z$.  On the other hand, $\delta_A$  is solely determined by the first term of \eref{fprimefstarEOM} and therefore has no cancellation between different contributions.  
 
 From Fig.~\ref{fig:imffstar} we can understand qualitatively why the asymmetry in scenarios \textbf{B} is suppressed with respect to $\textbf{A}$. However, from the differences in the CPV source one would expect a larger difference in baryon asymmetry than the factor $4$ we found and plotted in Fig.~\ref{BAUcutoff}. Not only is the source in scenario \textbf{A} five to ten times larger over the whole range of relevant $z$-values, but there is also no zero-crossing and therefore no associated cancellation between different contributions. So what causes the relatively small difference in baryon asymmetry between scenarios  \textbf{A} and  \textbf{B} compared to the much larger difference in the CPV source?

The baryon asymmetry is produced as the electroweak sphaleron process converts the produced chiral asymmetry into a baryon asymmetry in the symmetric phase in front of the bubble wall.  This chiral asymmetry $n_L$ is calculated from the transport equations.  In the symmetric phase the only non-zero rates on the r.h.s. of \eref{cascade} are the Yukawa and strong sphaleron interactions.  Deep inside the symmetric phase these interactions are (approximately) in equilibrium, and consequently the combination of number densities
\be
\left({2Q\over k_Q} -{R\over k_{R}}+{(Q+R)\over
    k_{\rm eff}}\right)\simeq 0 \,,\quad
\left({R\over k_{R}}-{H\over k_H}-{Q\over k_Q}\right) \simeq 0 \,,\qquad
(z < z_{\rm eq})
\ee
approximately vanish. Numerically, we find that this is an excellent approximation for $z_{\rm eq} \simeq -2\, \text{GeV}^{-1}$.  With these relations the chiral asymmetry can be written as a  $n_L= 5Q+4R = r_1 Q$, with
\be
r_1 =5 + \frac{4(2k_{\rm eff} +k_Q) k_R}{k_Q(k_{\rm eff} -k_R)}
\qquad (z < z_{\rm eq}). \label{r1}
\ee
The coefficient $r_1=0$ vanishes exactly in the massless limit, but also for our benchmark value it is small $r_1 \simeq -2 \times 10^{-2}$. The chiral asymmetry is thus much smaller than the individual number densities $n_L \ll H,Q,R$, as was also noted in Refs.~\cite{Tulin:2011wi,Chung:2008aya}.  

\begin{figure}
\begin{center}
\includegraphics[scale=0.62]{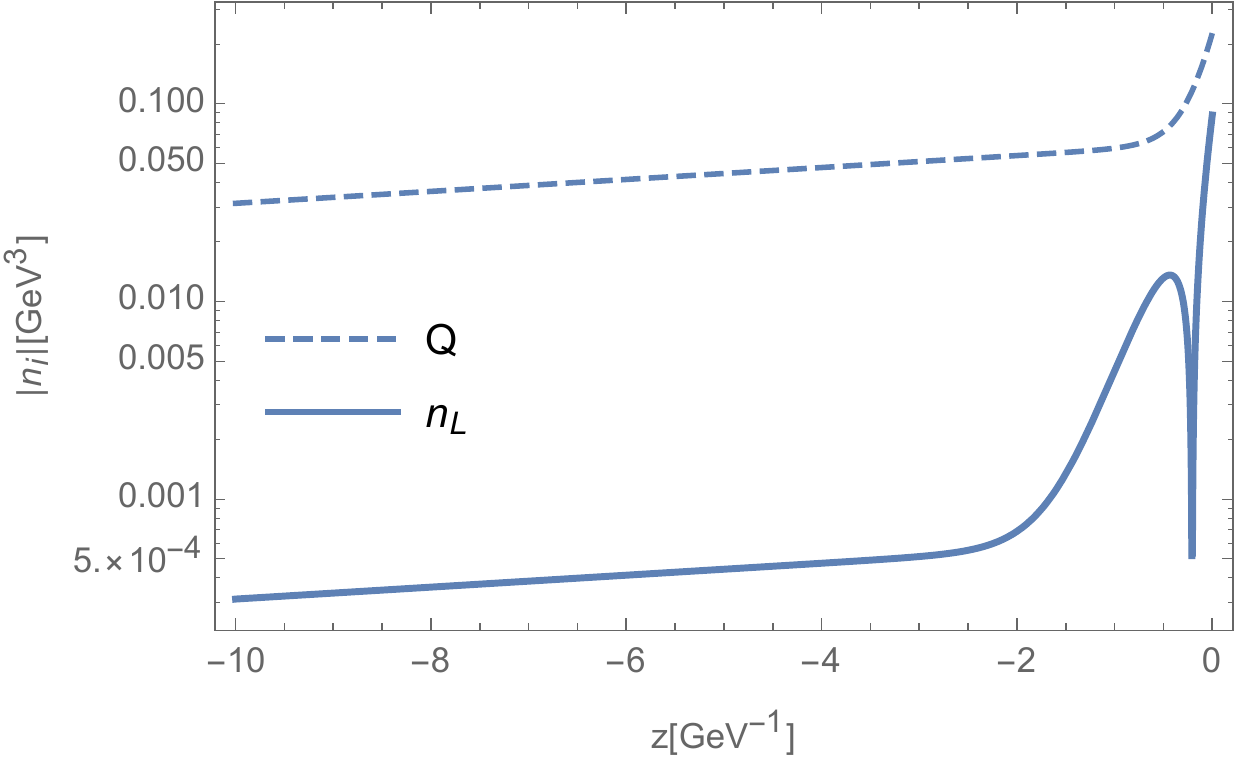}
\includegraphics[scale=0.62]{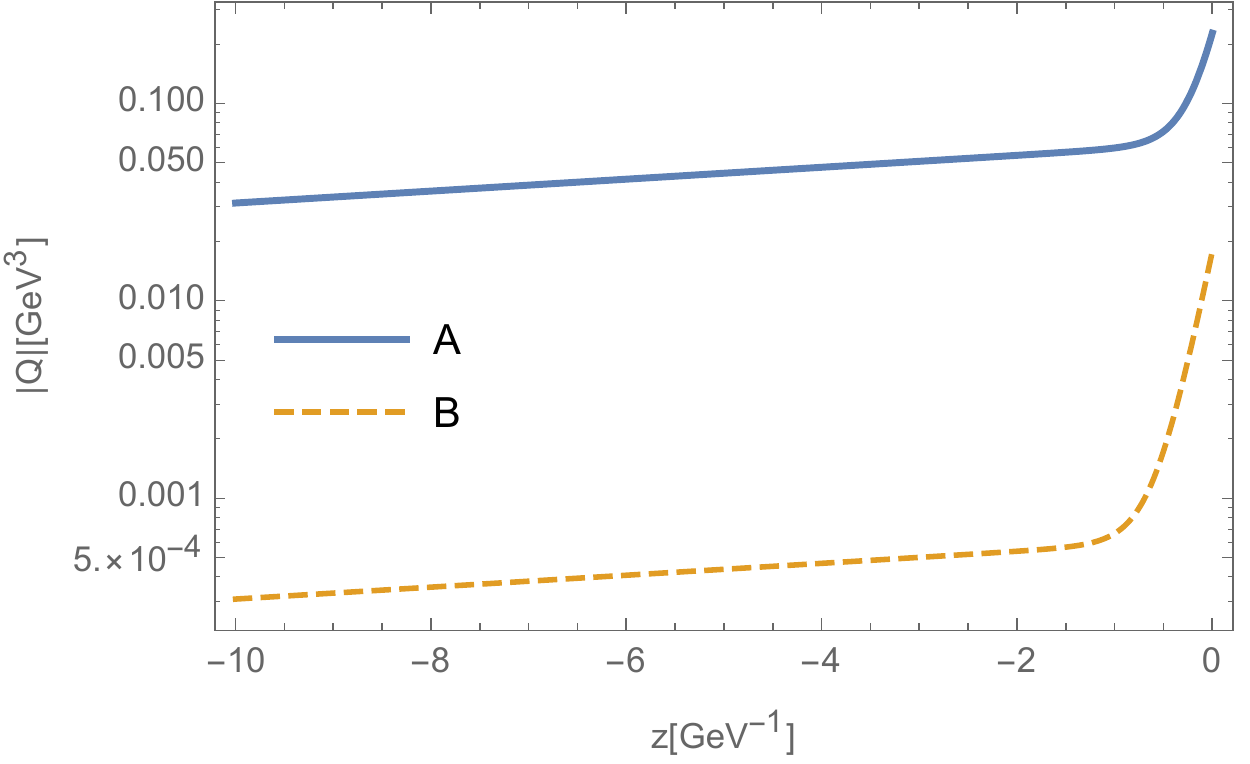}
\end{center}
\caption{Left panel: absolute value of the number density $Q$ (dashed blue) and $n_L$ (solid blue) in the symmetric phase for scenario \textbf{A}. The suppression of $n_L$ is especially efficient for $z<-2$. Right panel: absolute value of the number density $Q$ in the symmetric phase for scenario \textbf{A} (solid blue) and scenario \textbf{B} (dashed yellow). The density in scenario \textbf{A} is not only larger at the bubble wall at $z=0$, but it has also diffused into the symmetric phase more effectively. The number densities $T$ and $H$ show similar behavior.} \label{QandNLA}
\end{figure}

The physical picture is as follows. The source term is non-zero inside the bubble and creates a chiral asymmetry, which then diffuses into the symmetric phase. However, on scales far away from the bubble $z < z_{\rm eq}$, the strong sphaleron and Yukawa transitions are in equilibrium and suppress the chiral asymmetry. This is shown in the left panel of Fig.~\ref{QandNLA}, where the number density $n_L$ is compared with $Q$ in scenario \textbf{A}. Electroweak sphaleron transitions transform the chiral asymmetry into a baryon asymmetry.  However, this process is not efficient as only the small region right in front of the bubble $z_{\rm eq} < z < 0$ contributes significantly to the integral in \eref{nb_sol}. 

The above discussion is valid for both scenario {\textbf A}  and {\textbf B}, which only differ by the source term that vanishes in the symmetric phase.  It thus explains why in both scenarios baryogenesis is inefficient, and it is hard to obtain the observed asymmetry for cut-off scales consistent with EDM experiments.  The difference, however, is that the source is much larger in  scenario {\textbf A}.  Consequently, the non-zero number densities diffuse into the symmetric phase more efficiently, as becomes clear from the right panel of Fig.~\ref{QandNLA}, where the number density $Q$ is shown in the symmetric phase. If the cancellation in \eref{r1} is not taken into account\footnote{In fact, in the first arXiv version of this paper, we wrongly used the value 
\be k_{\rm eff} =\(\frac{2}{k_{Q_{1L}}}+
   \frac{2}{k_{Q_{2L}}}+\frac{1}{k_{U_R}}+\frac{1}{k_{C_R}}+\frac{1}{k_{D_R}}+\frac{1}{k_{S_R}}+\frac{1}{k_{B}}\)^{-1}
\ee
instead of \eref{keff}.  Although this changes $k_{\rm eff}$ by only approximately 20\%, the resulting $r_1 \simeq -0.5$ is much larger.  Consequently, with the wrong $k_{\rm eff} $ value  the whole region $z < z_{\rm eq}$ contributes in scenario {\rm A}, and the asymmetry is $\O(10^2)$ times larger than the  correct value.  In scenario {\rm B}, the wrong value for  $k_{\rm eff}$ only changes the result by $\O(1)$ effects, the reason being that the asymmetry is dominated by the region very close to the bubble wall in the first place.
}, and the  chiral asymmetry would be estimated by $n_L \sim O(Q)$ 
 the obtained asymmetry would be much larger, as now the whole region $z <z_{\rm eq}$ contributes (and actually gives the dominant contribution).  
 
 The source in scenario {\textbf B} is much smaller, and consequently there is less diffusion of number densities into the symmetric phase. In effect, the number densities are peaked very close to the bubble wall and the wash-out of $n_L$ due to the strong sphaleron and Yukawa interactions has a much smaller impact than for scenario \textbf{A}.   This explains why the large difference in the CPV sources between scenarios \textbf{A} and \textbf{B} as shown in Fig.~\ref{fig:imffstar}, are not completely transferred to large differences in the baryon asymmetry. 
 



\section{Discussion and conclusions}\label{conclusions} 
EWBG has been studied in many specific beyond-the-SM models. In this work, we have studied whether the crucial ingredients of EWBG can be studied without resorting to UV details of such models but instead by using effective operators to describe the EWPT and additional CP-violating sources. If applicable this would allow for a simple and model-independent description of a large class of models. Furthermore, specific SM extensions could be analyzed by matching to the EFT operators at the high-energy matching scale. As the SM-EFT operators can be and have been readily connected to low- and high-energy experiments this would allow for relatively easy tests of specific EWBG models. The main goal of this work was to study the effectiveness of the SM-EFT framework for EWBG. 

The premise of the SM-EFT framework is that operators can be ordered by their dimension with higher-dimensional operators giving rise to suppressed contributions with respect to lower-dimensional ones. Based on this premise, it is possible to derive a minimal basis of operators at a given order in the EFT expansion, see for instance Refs.~\cite{Buchmuller:1985jz, Grzadkowski:2010es}, by applying EOMs. Certain operators are then redundant, up to higher-order corrections, and can be eliminated. In particular, the CPV operator in scenario \textbf{B} is usually removed from the basis. A full EFT analysis would then include all relevant dimension-six operators in a minimal basis.  In this work we found that, for purposes of EWBG, the CPV operators in scenario \textbf{A} and \textbf{B} related by EOMs are not identical at all. The obtained baryon asymmetry differs by a large amount due to corrections from dimension-eight operators of the form
\begin{equation}
\mathcal L_8  =  C_{8}\,\bar Q_L\,\tilde \varphi\, t_R\,(\varphi^\dagger \varphi)^2\, ,
\label{L8}
\end{equation}
which therefore should be included in the analysis. 
In a general model-independent EFT approach there is then no, a priori, reason to not consider other dimension-eight operators that can contribute to the generation of the baryon asymmetry.  The starting assumption of the SM-EFT approach is thus explicitly violated, and it is not possible to study EWBG and the related phenomenology in a fully model-independent way.

The breakdown of the effective field theory might  be somewhat unexpected in view of the values of the scale of new physics required for successful baryogenesis. The scale corresponding to $\kappa = 2 \, \text{TeV}^{-2}\,$ is $\Lambda = 0.71 \,$ TeV while EDM experiments constrain $\Lambda_{CP} > 2.5$ TeV. There is no problem associated with the expansion in $\Lambda_{CP}$. While $\Lambda$ is relatively low, it is still significantly larger than all other physical scales in the computation of the baryon asymmetry and a perturbative expansion might seem reasonable. However, in order for a first-order EWPT to occur, the $(\varphi^\dagger \varphi)^3$-term needs to strongly modify the scalar potential as the SM itself is not capable of providing such a phase transition. The parameters are thus chosen such that the dimension-four and the dimension-six terms at the minimum are approximately equal during the phase transition. We have calculated how this lack of hierarchy between dimension-four and -six contributions in the scalar sector is transferred to the CPV sector. We have shown that this leads to no problems for the low-energy EDM phenomenology which is, to a large extent, identical in the two scenarios.  However, the CP-violating source which drives the generation of the baryon asymmetry is very different in the two scenarios leading to order-of-magnitude differences in particle number densities. Due to SM processes this large difference is not fully transferred to different baryon asymmetries, but nevertheless the total baryon asymmetry differs by a factor 4 in the two scenarios.



The difference in the baryon asymmetry raises the question whether EFT methods can still be useful for the study of EWBG.  The breakdown of the EFT approach originates from the scalar sector, while the CPV sector is in principle better under control. One potential approach is then to consider a concrete UV-complete model for the phase transition, but keep the EFT approach for the CPV sector. For example, a modification of the Higgs sector that has been studied extensively in the literature \cite{Espinosa:1993bs,Espinosa:2007qk,Barger:2007im,Espinosa:2008kw, Espinosa:2011ax,Cline:2012hg} is the addition of a $\mathbb{Z}_2$-symmetric singlet $S$. A UV-completion has the advantage that there is no expansion in the problematic scale $\Lambda$, and the redundancy between the operators is maintained, but the price to pay is that the description of the electroweak phase transition is no longer model-independent.  In addition, the EFT of the CPV sector needs to be extended to include effective operators that include the new singlet field. For a $\mathbb{Z}_2$-symmetric scenario, the first relevant operator is of the form $\mathcal L \sim  \bar Q_L \tilde \phi t_R |S|^{2}$ \cite{Vaskonen:2016yiu}. If the scalar field obtains a non-zero field value during the phase transition, these operators can give rise to a CPV phase contributing to EWBG that is not significantly constrained by EDM experiments. The direct link beween the baryon asymmetry and EDM experiments, which was present in the pure SM-EFT, is lost. 

Another proposal, put forward in Ref.~\cite{Balazs:2016yvi}, is to work with the full set of dimension-six CPV operators, and to not use the EOMs to remove redundancies.  In the context of the study of this paper, this means to treat the operators in scenario \textbf{A} and \textbf{B} as independent. Using a phenomenological description of the bubble wall in terms of a tanh-profile as in \eref{tanh}, the Higgs sector can be specified by a few parameters, and thus kept generic\footnote{For CPV derivative operators, the tanh-profile diverges in the center of the bubble, see the discussion below \eref{tanh}. This divergence may be tamed by a suitable regulator, but without knowing the exact Higgs potential and bubble profile, it is hard to estimate the error in this approximation.}. The disadvantage of this approach is that the physics leading to the phase transition can not be directly linked to collider experiments. Another problem is that the EDM or other low-energy experiments do not give constraints on the full set of dimension-six operators, since the redundancy is not broken at zero temperature such that observables only depend on a specific combination of operators that cannot be disentangled, even in principle. Finally, it is not clear how to match the EFT including redundant operators to a specific UV-complete model by an on-shell matching calculation at the high-energy scale.

In summary, we have investigated electroweak baryogenesis in the framework of the Standard Model EFT. We find that the EFT expansion breaks down due to the requirement of a strong first-order electroweak phase transition. We have shown that this also affects the expansion in the CP-violating sector of the EFT, such that higher-dimensional CP-violating operators cannot be a priori neglected. The pure Standard Model EFT is therefore not a suitable framework for electroweak baryogenesis. An extension of the EFT framework with additional scalar fields  but effective CPV operators might be more suitable at the cost of losing  model independence and a direct link to EDM phenomenology.


\section*{Acknowledgements}
We thank the Lorentz Center at the University of Leiden for its hospitality and support during the Snellius workshop ``Matter over Antimatter: The Sakharov Conditions After 50 Years" where this work was initiated.  We gratefully thank Bira van Kolck, Jason Yue, and Wouter Dekens for valuable discussions. We thank Ken Olum for help with the AnyBubble package. JdV is supported by the Dutch Organization for Scientific Research (NWO) through a VENI grant; MP and JvdV are supported by the research program of the Foundation for Fundamental Research on Matter (FOM), which is part of the Netherlands Organization for Scientific Research (NWO); and GW by the National Research Council of Canada.  This work was in part supported by the ARC Centre of Excellence for Particle Physics at the Terascale.

\appendix

\renewcommand{\theequation}{\thesection.\arabic{equation}}
\numberwithin{equation}{section}

\section{Thermal corrections}\label{A:thermal}

In this appendix we calculate the relevant thermal corrections.  We review the derivation of the finite temperature Higgs potential given in Ref.~\cite{Quiros:1999jp}, and then calculate the temperature corrections to the CPV couplings in scenario \textbf{A} and \textbf{B}.

The calculations are performed using the imaginary-time formalism, and in Landau gauge. Propagator for a scalar $\phi$, gauge boson $A_\mu$, and fermion $\psi$ field are given by
\begin{align}
G_\phi(k, \omega_n) &=-i \frac{1}{(\omega_n^2+ \vec k^2+ m_\phi^2)}=- i \Delta_\phi(k,\omega_n) \,, \nn\\
G_A^{\mu\nu}(k, \omega_n) &=i \frac{\(g^{\mu\nu}-\frac{k^{\mu}k^{\nu}}{m_A^2}\)}{(\omega_n^2+ \vec k^2+ m_A^2)}= i \(g^{\mu\nu}-\frac{k^{\mu}k^{\nu}}{m_A^2}\) \Delta_A(k,\omega_n) \,, \nn\\
S_\psi(k,\omega_n) &= -i\frac{ (\slashed{k}+m_\psi) }{(\omega_n^2+ \vec k^2+
                  m_f^2)} =- i(\slashed{k}+m_\psi) \Delta_f(k,\omega_n) \,,
\end{align}
with $k^\mu =(i\omega_n,\vec k)$ with $\omega_n = 2n\pi  / \beta$ for bosons, and $\omega_n = (2n+1)\pi / \beta$ for fermions. Here $\beta =1/T$ is the inverse temperature.  To complete the Feynman rules, for each loop and vertex interaction there is a factor
\be
{\rm loop} \; {\rm integral}: \;\; \frac{i}{\beta} \sum_{n=-\infty}^\infty \int \dbar^3 k\,,
\qquad \qquad
{\rm vertex}: \;\; -i \beta (2\pi)^3 \delta\(\sum_i \omega_i\) \delta^3\(\sum_i \vec k_i\)\,.
\ee
where we introduced the short-hand notation $\dbar k = \dd k/(2\pi)$.  We work in the hard thermal limit, which assumes that external momenta in Feynman diagrams are small compared to the temperature, implying, for example, that there are no thermal corrections to the kinetic terms.  Daisy diagrams are neglected.  The one-loop corrections can be separated into a zero- and a finite-temperature contribution.  The zero-temperature part is dealt with in the usual way by adding counterterms to absorb the divergencies, such that the effect of loop corrections can be captured by running coupling constants.  Here we focus on the temperature corrections.

 
\subsection{Higgs potential}

\begin{figure}
\begin{center}
\includegraphics[scale=.9]{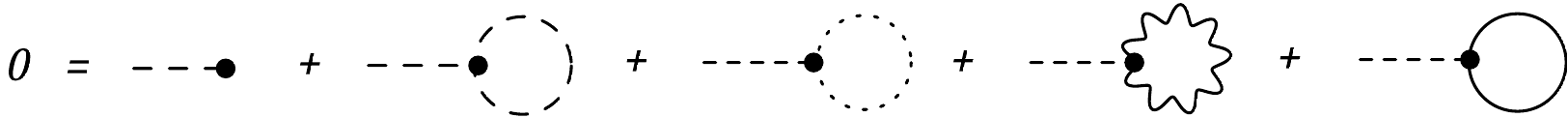}
\end{center}
\caption{Diagrammatic illustration of the vanishing of the Higgs tadpole at one-loop order. Solid lines denote top quarks, dashed lines the higgs boson, dotted lines the Goldstone bosons, and wavy lines the gauge bosons. Dots denote SM vertices. 
}\label{Fig:tadpole}
\end{figure}

Rather than calculate the potential, it is convenient to first calculate
the finite temperature equations of motion for the background Higgs
field $\phi_0$ \cite{Quiros:1999jp}.  The tadpole diagram is proportional to the equations of
motion, as the linear term in the Lagrangian vanishes on shell.
Explicitly
\be
\langle h \rangle = \Box \phi_0 +\frac{ \partial (V_0 + V_{\rm loop})}{\partial
  \phi_0} =0\,,
\label{tadpole}
\ee
with $V_0$ the tree-level potential in \eref{V0} and $V_{\rm loop}$ the one-loop contribution.  This relation is shown diagrammatically in Fig.~\ref{Fig:tadpole}. 
We write the interaction Lagrangian between $h$ and the other bosonic fields as
\begin{equation}
\mathcal L_{hXX}  =  \sum_{X=\theta_i,A,Z}\(- \frac{1}{2}\lambda_{hXX} h X^2 -  \frac{1}{3!}\lambda_{hhh} h^3\) \,,
\end{equation}
such that the loop contribution becomes
\begin{align}
\frac{\partial V_{\rm loop}}{\partial \phi_0} &= \sum_{X=h,\theta_i,A,Z}  n_X \frac12 \lambda_{hXX}
\frac1{\beta} \sum_n \int \dbar^3 p\, \Delta_X(\omega_n,p) - 
y_t  \frac1{\beta} \sum_n \int \dbar^3 p \, \Tr
S_t(\omega_n,p) 
\label{dVloop}
 \\
&= \sum_{X=h,\theta_i,A,Z}  n_X \frac12 \partial_\phi m_X^2
\frac1{\beta} \sum_n \int \dbar^3 p\, \Delta_X(\omega_n,p) - n_t
(\frac12 \partial_\phi m_t^2) \frac1{\beta} \sum_n \int \dbar^3 p \,
\Delta_t(\omega_n,p) 
\nn
\end{align}
with $n_{\{h,\theta,W,Z, t\}} = \{1, 3, 6,3,4N_C\}$ and we used the short hand notation $\dbar k = \dd k/(2\pi)$. The first term gives the contributions of the bosonic loops (the higgs field, the three goldstone bosons, the $W^\pm$ and $Z$ fields) and the second term of the fermionic loops where we only included the top quark.  The factor $1/2$ in the bosonic terms is a symmetry factor.  %
To get the second expression in \eref{dVloop} we used that the the bosonic trilinear couplings can be written as $\lambda_{hXX} = (\partial m_X^2/\partial \phi_0) \equiv \partial_\phi m_X^2$, and similarly the Yukawa coupling $y_f = \partial_\phi m_f$. Finally, the trace of the fermion propagator gives a factor $4N_c m_f$ for a colored Dirac fermion.

The contribution from a particular boson $X$ can be written as
\be
\frac{\partial V^{(X)}_{\rm loop}}{\partial m_X^2} = \frac1{2\beta} \sum_n \int \dbar^3
p\, \Delta_X(\omega_n,p)
= \frac12 \int \dbar^3
p\, \( \frac1{2\omega} + \frac1\omega \frac{1}{\e^{\beta \omega}-1} \)\,.
\ee
 This expression can be integrated to give
\begin{align}
V^{(X)}_{\rm loop} &= \frac12 \int \dbar^4
p\, \ln(p^2+m_X^2) + \frac1{2\pi^2\beta^4} J_B (m_X^2 \beta^2) 
\equiv V^{(X)}_1 + V^{(X)}_T\,,
\end{align}
with $V^{(X)}_1$ the zero-temperature one-loop result --- the
Coleman-Weinberg potential --- and $V^{(X)}_T$ the finite-temperature
contribution. 

The same steps can be followed for the contribution from a fermion
field; here we concentrate on the top contribution
\be
\frac{\partial V^{(t)}_{\rm loop}}{\partial m_t^2} = - \frac{4N_c}{2\beta} \sum_n \int \dbar^3
p\, \Delta_t(\omega_n,p)
= -2 N_c \int \dbar^3
p\, \( \frac1{2\omega} - \frac1\omega \frac{1}{\e^{\beta \omega}+1} \)\,.
\ee
Integrating this expression gives
\be
V^{(t)}_{\rm loop} = -2N_c \frac12 \int \dbar^4
p\, \ln(p^2+m_t^2) -4N_c \frac1{2\pi^2\beta^4} J_F (m_t^2 \beta^2)
\equiv  V_1^{(t)} + V^{(t)}_T \,.
\ee

Adding up the contributions of all fields, the 1-loop
finite-temperature potential becomes
\begin{align}\label{VT2}
V_T &=\sum_X n_X \frac{T^4}{2\pi^2} J_B (m_X^2/T^2)
-  n_t \frac{T^4}{2\pi^2} J_F (m_t^2/T^2) \nn \\
 &= \sum_X n_X \frac{1}{24} m_X^2 T^2 
+  n_t \frac1{48} m_t^2 T^2 +\O(T)\,,
\end{align}
where the second line gives the leading field-dependent terms in the high-temperature expansion, for which we used
\be
J_B(x^2) = -\frac{\pi^4}{ 45}+ \frac{\pi^2}{12} x^2 - \frac{\pi}{6} x^3
+ \cdots\,, \qquad
J_F(x^2) = -\frac{7\pi^4}{360} -\frac{\pi^2}{24} x^2 + \cdots\,.
\label{Jexpand}
\ee
The validity of the high temperature expansion can be improved by including the cubic $T^3$ term.  For consistency, one then also needs to include the daisy diagrams which contribute at the same order.
Since we neglect the latter, we also neglect the cubic term in our analysis.  Finally, for future use, we list the following relations for bosons and fermions respectively
\begin{align}
n_X \left(\frac12 \partial_\phi m_X^2\right)
\frac1{\beta} \sum_n \int \dbar^3 p\, \Delta_X(\omega_n,p) &= 
\frac{\partial}{\partial \phi_0} \[V_1^{(X)} + V_T^{(X)}\]
\,,
\nn \\
n_t\left(\frac12 \partial_\phi m_t^2\right) \frac1{\beta} \sum_n \int \dbar^3 p \,
\Delta_t(\omega_n,p)  &= \frac{\partial}{\partial \phi_0} \[V_1^{(t)} + V_T^{(t)}\]\,.
\label{dVrelation}
\end{align}
%


\subsection{Thermal corrections to the CP-violating source term}\label{A:f}

We now turn to the CPV sector. We consider the scenarios given in Sect.~\ref{EFT} with the following CPV operators relevant for baryogenesis
\begin{eqnarray}
\mathcal L^{(A)}_{\mathrm{CPV}} &=& - C_Y\, \bar Q_L  y_t \tilde \varphi \,t_R\,(\varphi^\dagger \varphi)+ \textrm{h.c.}\,,\nonumber \\
\mathcal L^{(B)}_{\mathrm{CPV}} &=&  - \alpha C_{DD}\, \bar Q_L D^2\tilde \varphi \,t_R\, + \textrm{h.c.}\,,\nonumber \\
\mathcal L^{(\mathrm{EOM})}_{\mathrm{CPV}} &=& -\alpha C_{DD} \bar Q_L  \tilde\varphi \,t_R\, \frac{\mathcal L_{\varphi}}{\partial (\varphi^\dagger \varphi)}+ \textrm{h.c.}\,,\label{CPVA}
\end{eqnarray}
where in the last scenario $\mathcal L_\varphi = -\mu^2 (\varphi^\dagger \varphi) -\lambda (\varphi^\dagger \varphi)^2 - \kappa  (\varphi^\dagger \varphi)^3$ corresponding to the tree-level potential.
We will first calculate the thermal corrections to the CPV mass terms in   $\mathcal L^{(B)}_{\mathrm{CPV}}$ and $\mathcal L^{(\mathrm{EOM})}_{\mathrm{CPV}}$, since those for scenario \textbf{A} then follow from the relation $\mathcal L^{(A)}_{\mathrm{CPV}} = \mathcal L^{(\mathrm{EOM})}_{\mathrm{CPV}} |_{\mu^2=\kappa=0}$ and replacing $C_{DD}\rightarrow -(y_t C_Y)/(2\alpha\lambda)$. 

\begin{figure}
\begin{center}
\includegraphics[scale=.9]{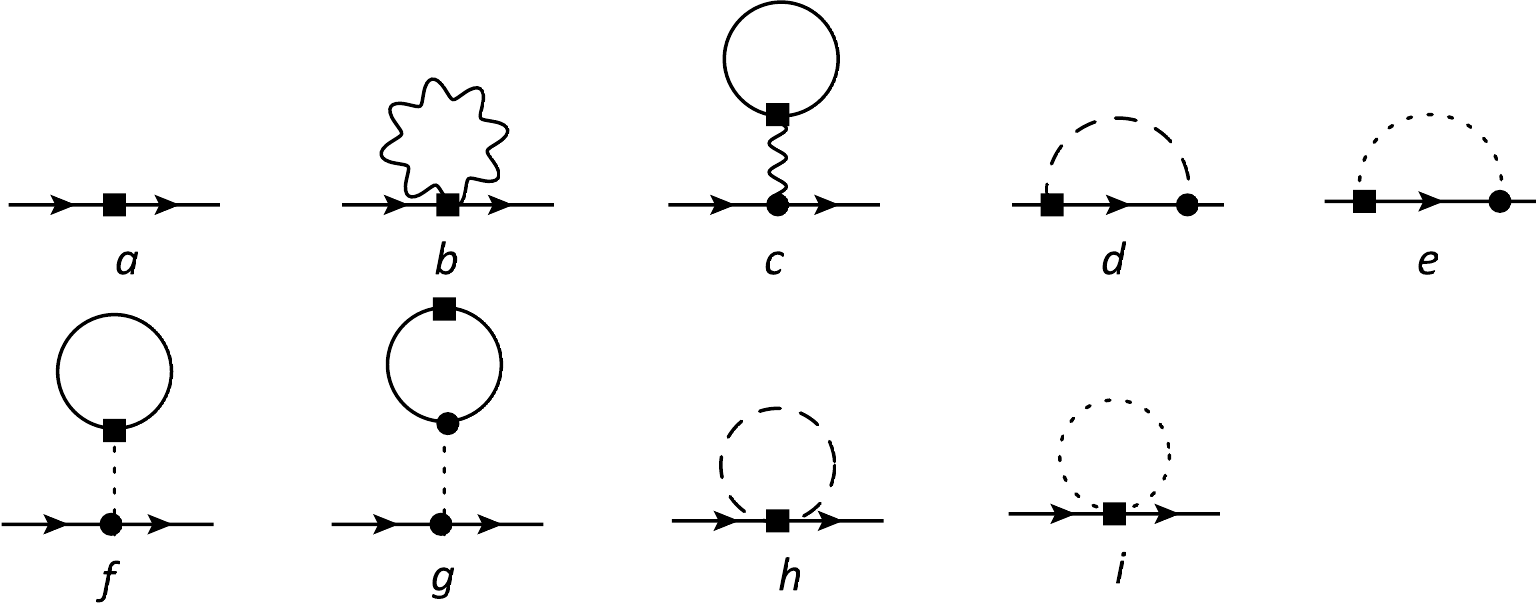}
\end{center}
\caption{Tree- and one-loop diagrams contribution to the CP-violating source term in the three scenarios under investigation. Squares denote interactions from Eq.~\eqref{CPVA}. For the other notation, see Fig.~\ref{Fig:tadpole}. Only one topology for each diagram is shown. 
}\label{scenB}
\end{figure}

We first consider scenario \textbf{B}. The diagrams contributing to thermal corrections to the CPV source are shown in Fig.~\ref{scenB}. Diagram $a$ is the tree-level contribution proportional to $\Box \phi_0$, which would vanish in the zero-temperature minimum. Diagram $b$ and $c$ appear from the gauge-dependent parts of the covariant derivates in the CPV operator in scenario \textbf{B}. Diagram $c$ vanishes, however, and diagram $b$ is discussed below.  It turns out that the finite-temperature contributions of diagrams $d$ and $e$ vanish in the hard thermal limit, which we apply here, and we neglect them. Diagram $f$ vanishes because there is no momentum flowing in the CPV vertex. Diagram $g$ is also proportional to $\Box \phi_0$ which, as it appears in a one-loop diagram, we trade for the derivative of the tree-level potential, $-\partial V_0/\partial \phi_0 = -m^2_\theta \phi_0$, where  $m_\theta $ is the mass of the goldstone bosons (this mass vanishes of course in the zero-temperature minimum). The goldstone mass cancels against the zero-momentum goldstone propagator appearing in diagram $g$. Finally, diagrams $h$ and $i$ are not present in scenario $\textbf{B}$. The non-vanishing diagrams sum to
\begin{align}
 \A^{(B)} &= \bar t (i\gamma^5) t  \, \left(\frac{-\alpha \tilde c_{DD}}{\sqrt{2}}\right)
\bigg[ \Box \phi_0 \nn \\
&+ \sum_{X=W,Z} n_X \frac12\partial_\phi
m_X^2 \frac1{\beta} \sum \int
\dbar^3 k\Delta_X(k,\omega_n) - n_t(\frac{1}{2}\partial_\phi m_t^2)\frac{1}{\beta}\sum_n\int \dbar^3
p\, \Delta_t(\omega_n,p)
\bigg] \nn \\
&=  \bar t (i\gamma^5) t  \, \left(\frac{-\alpha \tilde c_{DD}}{\sqrt{2}}\right)\[\Box \phi_0
+\frac{\partial
  (V^{(W,Z)}_1+ V^{(W,Z)}_T+ V^{(t)}_1+ V^{(t)}_T)}{\partial \phi_0}\]\,,
  \label{AB}
\end{align}
where $V_1^{(W,Z,t)}$ and  $V_T^{(W,Z,t)}$ denote, respectively, the zero- and finite-temperature
contributions to the one-loop effective potential from the gauge
bosons and the top quark.  In the last equality we applied Equations  (\ref{dVrelation}) and (\ref{VT2}).

We now turn to $\mathcal L^{(\mathrm{EOM})}_{\mathrm{CPV}}$ and calculate the  correction to the CPV source terms. Again the relevant diagrams are depicted in Fig.~\ref{scenB}. Diagram $a$ is now proportional to $\partial V_0/\partial \phi_0$ as can be seen from Eq.~\eqref{CPVA}. Diagrams $b$ and $c$ are not relevant as the corresponding CPV vertices are not present, and diagrams $d$ and $e$ vanish again in the hard thermal limit. Diagrams $f$ and $g$ are both nonzero, but they mutually cancel. Diagrams $h$ and $i$ are nonzero and together with $a$, we obtain
\begin{align}
 \A^{(\mathrm{EOM})} &=  \bar t (i\gamma^5) t  \, \frac{\alpha \tilde c_{DD}}{\sqrt{2}} \bigg[
\frac{\partial V_0}{\partial \phi_0} 
+\sum_{X=h,\theta} n_X\frac12\partial_\phi
m_X^2 \frac1{\beta} \sum \int
\dbar^3 k\Delta_X(k,\omega_n)\bigg]
  \\
&=  \bar t (i\gamma^5) t  \, \frac{\alpha \tilde c_{DD}}{\sqrt{2}} \[\frac{\partial
  (V_0+V^{(h,\theta)}_1+ V^ {(h,\theta)}_T)}{\partial \phi_0}\]\,,
\label{AR}
\end{align}
where we again applied \eref{dVrelation}. A comparison of Eqs. \eqref{AB} and \eqref{AR} now shows that $\A^{(B)}$ and $\A^{(\mathrm{EOM})}$ are related by the one-loop corrected EOM, $\Box \phi_0 = - \partial (V_0+V_1+V_T)/\partial \phi_0 =- \partial_\phi V$, as expected. 

For $\mathcal L^{(A)}_{\mathrm{CPV}}$, the same diagrams contribute as for $\mathcal L^{(\mathrm{EOM})}_{\mathrm{CPV}}$. That is, only diagrams $a$, $h$, and $i$ are nonzero.  We have to be a bit more
careful now, as the CPV vertices appearing in $h$ and $i$ are now proportional to $\partial_\phi m_X^2 |_{\mu^2=\kappa=0}$, while the masses of the Higgs and goldstone bosons in
the propagators still depend on the full tree-level potential.  The result therefore becomes
\begin{align}
 \A^{(A)} &=  \bar t (i\gamma^5) t  \, \left(\frac{-\tilde c_Y y_t}{2\sqrt{2}\lambda}\right) \[
\frac{\partial V_0}{\partial \phi_0}\big|_{\mu=\kappa=0}+ \sum_{X=h,\theta}
       n_X\frac12 (\partial_\phi
m_X^2)_{\mu=\kappa=0}\frac1{\beta} \sum \int
\dbar^3 k\Delta_X(k,\omega_n)\] \nn
  \\
&=  \bar t (i\gamma^5) t  \,  \left(\frac{-\tilde c_Y y_t}{2\sqrt{2}\lambda}\right)\[
 \frac{\partial V_0}{\partial \phi_0}\big|_{\mu=\kappa=0}+\sum_{h,\theta} \frac{(\partial_\phi
  m_X^2)_{\mu=\kappa=0}}{(\partial_\phi m_X^2)}
\partial_\phi  (V^{X}_1+ V^{X}_T)\]\,.
\end{align}

We are now in the position to calculate the interaction terms defined in \eref{Lint} 
\begin{align}
f_A(\phi_b) &=( y_t\phi_b + \frac{i}{2}y_t\tilde c_{Y} \phi_b^3) + i\frac{\tilde c_Y y_t}{2\lambda}\[\sum_{h,\theta} \frac{(\partial_\phi
  m_X^2)_{\mu=\kappa=0}}{(\partial_\phi m_X^2)} 
\frac{\partial  V^{X}_T}{\partial \phi_0}\]\bigg|_{\phi_0 = \phi_b}\,,\nn \\
f_B(\phi_b) &= (y_t \phi_b + i\alpha \tilde c_{DD} \Box \phi_b) +i\alpha \tilde c_{DD} \[\frac{\partial  (V^{(W,Z)}_T+V^{(t)}_T)}{\partial \phi_0} \]\bigg|_{\phi_0 = \phi_b} \,,\nn \\
f_\text{EOM}(\phi_b) &=y_t \phi_b - i\alpha \tilde c_{DD} \left( \frac{\partial V_0}{\partial \phi_0}+\frac{\partial  V^ {(h,\theta)}_T}{\partial \phi_0}\right)\bigg |_{\phi_0=\phi_b} \,. 
\label{ffunctions}
\end{align}
In the high-temperature limit we can easily take the derivatives 
\begin{equation}
\frac{\partial  V^{(X)}_T}{\partial \phi_0} = n_X\frac{T^2}{24 } \partial_\phi m_X^2\,,\qquad 
\frac{\partial  V^{(t)}_T}{\partial \phi_0} =  n_t \frac{T^2}{48} \partial_\phi m_t^2\,,
\end{equation}
from which we obtain
\begin{align}
f_A(\phi_b)& =  y_t \phi_b+\frac{iy_t \tilde c_Y}{2}\left[\frac{1}{2} T^2 \phi_b +\phi_b^3\right],\nn \\
f_B(\phi_b)&=y_t \phi_b+i\alpha \tilde c_{DD} \left[\Box \phi_b + \frac{T^2}{16}(3g^2 + g'^2+ 4 y_t^2)\phi_b \right]\,, \nn \\ 
f_\text{EOM}(\phi_b)&=y_t \phi_b-i\alpha \tilde c_{DD}\[\(\mu^2 + \frac{T^2}{2}\lambda \)\phi_b+ \(\lambda+ \kappa T^2\) \phi_b^3 + \frac{3\kappa}{4}\phi_b^5 \]\,.
\label{fnum}
\end{align}
where we should use \eref{mulambda} to substitute for $\mu^2$ and $\lambda$ in $f^{(\mathrm{EOM})}(\phi_b)$. Since we are in the rest frame of the bubble wall, the derivative operator $\Box$ reduces to the three-dimensional Laplace operator in spherical coordinates (with a minus sign due to the metric).


\section{Rates and parameters and transport equations}\label{A:input}

In appendix \ref{A:rates} we list all the rates and parameters that appear in the full set of transport equations (\ref{cascade}), which are needed to reproduce our results.  In appendix \ref{A:solution} we quickly review how to solve the transport equations semi-analytically. 

\subsection{Rates and parameters}\label{A:rates}

For the values of the coupling constants at the electroweak scale $\mu= m_Z$ we use
\be g' = 0.36\,, \quad  g = 0.65\,, \quad  g_s =1.23\,, \,\quad
y_t = 1\,,\quad y_b =0.1\,.
\ee
The diffusion coefficients are \cite{Cline:2000nw}
\be
D_T \simeq \frac6{T}\,,\quad D_Q \simeq \frac6{T}\,, \quad D_H \simeq
\frac{100}{T}\,.
\label{diffusion}
\ee
The SM thermal masses $\delta m_i^{\rm Re}$ are \cite{Enqvist:1997ff}
\begin{align}
(\delta m_Q^{\rm Re})^2 &= \(\frac16 g_s^2 +\frac{3}{32} g^2 +\frac1{216} g'^{2} +\frac1{16}
y_t^2\)T^2\,, \nn \\
(\delta m_{t_R} ^{\rm Re})^2&= \(\frac16 g_s^2 +\frac1{18} g'^{2} +\frac1{8}
y_t^2\)T^2\,,\nn\\
(\delta m_H^{\rm Re})^2&=\(\frac{3}{16}g^2
+\frac1{16} g'^{2}+\frac14 y_t^2 +\frac14 \(\frac{m_H^2}{v_0^2}-3v_0^2 \kappa\) + 3\phi_0^2 \kappa \)T^2\,.
\label{deltamR}
\end{align}
For the Higgs mass we included the contribution from the dimension six
operator $\kappa (\varphi^\dagger \varphi)^3$. The result for the thermal masses can be derived from the
effective potential in the high-temperature expansion \eref{VTexpand}.

The $k$-function can be written as
\begin{equation}
k_i(x) = k_i(0)\frac{c_{F,B}}{\pi^2}\int_{m/T}^\infty dx\,x\,
\frac{e^x}{(e^x \pm 1)^2}\sqrt{x^2 - m^2/T^2}\,,
\label{kfun}
\end{equation}
where $c_{F(B)} = 6\,(3)$ and the $+(-)$ sign in the denominator are
for fermions (bosons), and
\be
k_{Q_L}(0) = 6\,, \quad k_{Q_R}(0)= 3\,, \quad k_H(0) =4\,,
\label{k0}
\ee
where $Q_L$ denotes any left-handed quark doublet and $Q_R$ any right-handed quark singlet.
These functions are calculated for $x = (m_i^\Re +\delta m_i^\Re)/T$
and thus give slightly different values in the symmetric and broken
phase. 

The relaxation rate, source term, and Yukawa rate are all calculated neglecting collective
plasma excitations (hole modes). The relaxation rate and CPV source term are
\begin{align}
\Gamma_M^\pm &= \frac{6}{T^2} \times
\frac{N_c}{2\pi^2 T} | f|^2  
\int \frac{k^2 \dd k }{\omega_L
  \omega_R} \Im\bigg[ -
 \frac{\( h(\E_L) \mp  h(\E_R^*) \)}{\E_R^* -\E_L}\(
\E_L \E_R^* - k^2 \) \nn \\
& \hspace{3.5cm}
+\frac{\(h(\E_L) \mp  h(\E_R) \)}{\E_L +\E_R}\(
\E_L \E_R +k^2 \) \bigg]\,,
\end{align}
and
\begin{align}
S_R^{\CPV} &= 
\frac{v_w N_c}{\pi^2} \Im\(f' f^*\) 
  \int \frac{k^2 \dd k }{\omega_L
  \omega_R} \Im \bigg[ \nn
\frac{(n_f(\E_L) -n_f(\E_R^*))}{(\E_L -\E_R^*)^2}
\( \E_L \E_R^* -k^2\) \nn \\
&\hspace{4cm}
+ \frac{(n_f(\E_L) +n_f(\E_R)-1) }{(\E_L +\E_R)^2}
\( \E_L
 \E_R +k^2\)\bigg]\,,
\end{align}
with $|f|^2 = y_t^2 \phi_b^2 + \O(\Lambda_{CP}^{-4})$, $N_c =3$ the number of colors, $n_f(x) = (\e^x+1)^{-1}$ the Fermi-Dirac distribution, and $h_F (x) =\e^x (e^x +1)^{-2}$.  Further $\E_i = \omega_i - i \Gamma_t = \sqrt{k^2 + (\delta m_i^{\rm Re})^2} -i \Gamma_t$.\footnote{If
  the convention $\E_i = \omega_i + i \Gamma_t$ is chosen, this would give
  an overall sign difference for the rate $\Gamma_M^\pm$ and source
  $S_R^{\CPV}$.}  The thermal width is
$\Gamma_t \simeq g_s^2 T C_f/(4\pi) = 0.16\, T$ with $C_F =4/3$ \cite{Joyce:1994zn,Elmfors:1998hh}.

The Yukawa rate is $\Gamma_Y=\Gamma^{(3)}_Y+\Gamma^{(4)}_Y$, where the
3-point  $(\bar t_L t_R h)$-interaction is given by \cite{Cirigliano:2006wh}
\bea
\Gamma^{(3)}_Y &&= \frac{12 N_c y_t^2}{T^2}  \frac1{16\pi^3} (m_R^2 +m_Q^2
-m_H^2) \int_{m_T}^\infty \dd \omega_R \frac{\e^{\omega_R/T}}{(\e^{\omega_R/T} +
  1)^2}
\Bigg\{ \nn \\
&&
\ln\( \frac{ (\e^{\omega_H^+/T}-1)
  (\e^{\omega_H^-/T}+\e^{\omega_R/T})}
{ (\e^{\omega_H^-/T}-1) (\e^{\omega_H^+/T}+\e^{\omega_R/T})}\)
\[ \theta(m_R-m_Q-m_H)-\theta(m_H -m_R-m_Q)\] \nn \\
&&+\ln\( \frac{ (\e^{\omega_H^+/T}-1)
  (\e^{\omega_H^-/T}+\e^{-\omega_R/T})}
{ (\e^{\omega_H^-/T}-1) (\e^{\omega_H^+/T}+\e^{-\omega_R/T})}\)
\theta(m_Q-m_R-m_H) \Bigg\}\,,
\eea
with
\be
\begin{aligned}
\omega_H^\pm = &\frac1{2m_R^2} \Bigg\{ \omega_R|m_H^2 +m_R^2-m_L^2|  \\ &\pm 
\sqrt{(\omega_R^2 -m_R^2)(m_R^2-(m_L+m_H)^2) (m_R^2 -(m_L-m_H)^2)}
\Bigg\}\,.
\end{aligned}
\ee
In these expressions all mass parameters should be taken as $\delta
m_i^\Re$. For the parameters of interest this rate is kinematically
forbidden, and all the Heaviside functions in the above expression
vanish. For this reason we also include the 4-point $(\bar t_L t_R h
g)$-interaction, that is, with one extra gluon line \cite{Cirigliano:2006wh}
\be
\Gamma_Y^{(4)} = \frac{\zeta_3}{6\pi^2} g_s^2 |f|^2 T
\ln\(\frac{8T^2}{ m_q^2}\)\,,
\ee
 where we take for the typical mass scale $m_q = \delta m_t^{\rm Re}$.

The strong sphaleron rate is\cite{Moore:2010jd}
\be
\Gamma_{ss}  = 14\kappa' \alpha_{s}^4 T\,,\ee
where we take $\kappa'=1$.

\subsection{Solving the transport equations}\label{A:solution}

To solve the transport equations semi-analytically we use the method of Ref.~\cite{White:2015bva}, which we summarize here.  
We take the transport equations as written in \eref{cascade}. Neglecting the bubble wall curvature, the right-hand side of the equations simplifies to $\partial_\mu J_X^\mu  = v_w n_X' - D_X  n_X''$, and the 
equations are of the general form
\bea
a^i_{Q1} \partial^i Q+a^i_{R1} \partial^i R &=&0 \label{eq1}\,, \\
a^i_{Q2} \partial^i Q+a^i_{R2} \partial^i R+a^i_{H2} \partial^i H &=&0\,, \label{eq2} \\
a^i_{Q3} \partial^i Q+a^i_{R3} \partial^i R+a^i_{H3} \partial^i H &
=&\Delta(z)\,.
\label{eq3} 
\eea
The coefficients $a_{Xj}^i$ can be read off by comparing with
\eref{cascade}. The index $i=0,1,2$ labels the number of
derivatives, $j=1,2,3$ the equation number, and $X={Q,R,H}$ the
fields. The source is $\Delta = S_{R}^{\CPV} $.  Note that all
coefficients are just numbers once all input parameters and rates in
the transport equations have been specified.

The system of equations is solved in the symmetric ($z<0$) and in the
broken phase ($z>0$) separately, and then matched together at the
location of the bubble wall at $z=0$. We take a thin wall limit, where 
the coefficients $a_{Xj}^i$ are taken constant (but possibly different) in both
phases. The $z$-dependence of the source is taken into account. Since
the source is peaked mainly in the broken phase, to a good
approximation we can set $\Delta =0$ in the symmetric phase.

The first step is to solve the homogeneous equations, that is with the
right-hand-side set to zero in \eref{eq3}.  We use the Ansatz
$X = A_X \e^{\alpha z} \equiv A_X l(z)$, with $X=\{Q,R,H\}$.  Plugging the Ansatz
into \eref{eq1} and \eref{eq2} gives the first two expressions below:
\be
A_{R} = -A_Q \frac{a_{Q1}^i \alpha^i}{a_{R1}^j \alpha^j}\,, \quad
A_H = - \frac{A_Q a_{Q2}^i \alpha^i +A_{R} a_{R2}^i \alpha^i}{a_{H2}^j
  \alpha^j}\,, \quad
A_Q = (a_{R1}^i \alpha^i)(a_{H2}^j \alpha^j)\,,
\label{Aval}
\ee
where indices are summed over.  Note that the superscript on
$a_{Xj}^i$ is an index whereas on $\alpha^i$ it gives the power.  We
have further choosen $A_Q $ such that when plugged into \eref{eq3} there
are no denominators (the normalization of $A_Q$ is arbitrary, it will
give a different condition on the normalization constants $x_i$ defined
below, such that in the end the solution is the same):
\bea
0
&=& \( a_{R1}^i a_{H2}^j a_{Q3}^k +a_{Q1}^i a_{R2}^j
a_{H3}^k - a_{Q1}^i a_{H2}^j a_{R3}^k - a_{R1}^i a_{Q2}^j a_{H3}^k\) 
\delta_{i+j+k-n}\partial^n l(z) \nn \\ 
&\equiv&  a_n \partial^n l(z)  \qquad \Rightarrow \qquad a_n \alpha^n =0\,,
\label{root}
\eea
where there is a sum over repeated indices.  This equation is solved
in the broken and symmetric phase. It has six real
roots, 3 of which are positive and 3 negative (or two negative and one
zero). In the broken phase the roots are denoted by $\alpha_i$, in the symmetric phase by $\gamma_i$, with
$i=1,...,6$. We order them as follows
\bea
&\alpha_i \leq 0 \;\; {\rm for} \; i=1,2,3\ \qquad&\alpha_i > 0 \;\; {\rm
  for} \; i=4,5,6 \nn \\
&\gamma_i >0  \;\; {\rm for} \; i=1,2,3 \qquad &\gamma_i \leq 0 \;\; {\rm for} \; i=4,5,6\,.
\label{ordering}
\eea

As mentioned above, in the symmetric phase the source (approximately)
vanishes, and the solution to the homogeneous equations is  also
the full solution
\be
X = \sum_i A_X[\gamma_i] y_i \e^{\gamma_i z}, \quad ({\rm symmetric})\,.
\ee
Here $A_X[\gamma_i]$ is just a number, which is different for each root $\gamma_i$ and $y_i$ is an integration constant. 

In the broken phase the full solution satisfies the inhomogeneous
equation with $\Delta(z) \neq 0$.  We write $l(z) = \sum_i l_i$ and
$X = \sum_i A_X[\alpha_i] l_i $, where the sum is over the six roots, and
\be
l = \sum_i  x_i \e^{\alpha_i z} \( \int_0^z \e^{-\alpha_i y}
\Delta(y) \dd y - \beta_i \).
\label{solbroken}
\ee
The normalization constants $x_i$ are fixed by requiring the above
solution to be a solution of the inhomogeneous equations; plugging
\eref{solbroken} into Eqs.  (\ref{eq1}-\ref{eq3}), this is the case if
\be
\Delta(z) = \sum_{n=0}^6 a_n \partial^n l(z)  =
\sum_i \sum_{n \geq j \geq 1} a_n x_i
\alpha_i^{n-j} \partial_z^{(j-1)}\Delta(z)\,.
\label{fix_x}
\ee
Here we used that when all derivatives act on the $e^{\alpha z}$ part
in $l$ --- this is the ($j=0$)-term left out above --- this gives an
expression that is proportional to the root equation $a_n \alpha^n$,
and this vanishes.  The summation thus starts at $j=1$ and each term
comes from $j$ derivatives on the integral part in $l$.  Note that
there is only one order in which the derivatives can be taken, because
once the derivative of the integral in $X$ is taken the $z$-dependent
prefactor $\e^{\alpha_i z}$ cancels, and only further derivatives of
$\Delta$ are possible.  The coefficients of the different
$\partial_z^{(j-1)}\Delta $ contributions all vanish, except for the
$(j=1)$-contribution which should match the $\Delta(z)$ on the
left-hand-side of \eref{fix_x}.
Solving the $j=2,...,6$
equations gives the set of constraints $\sum x_i \alpha_i^k =0$ for
$k=0,...,4$. The $j=1$ equation gives $\sum x_i \alpha_i^5 = 1/a_6$.
This can be written in the short-hand form
\be
\sum_i x_i \alpha_i^j = \frac{\delta_{j5}}{a_6}\,, \qquad j=0,...,5\,.
\label{Xi}
\ee

We now found the solutions in the symmetric and broken phase. There
are 12 more integration constants $\beta_i$ and $y_i$.  Six of them are
fixed by the boundary conditions, and six by matching the solution in
a continuous way at $z=0$. The boundary conditions are that the number
densities vanish at infinity. For the symmetric phase the number
densities should vanish at $z = -\infty$. For $\gamma_i \leq 0$ this can
only be satisfied if
\be
y_i =0 \; \;\forall \; \gamma_i \leq 0 \;\; ({\rm for} \; i=4,5,6)\,.
\ee
In the broken phase the number densities should vanish at $z = \infty$. For
$\alpha_i>0$ this is satisfied if
\be
\beta_i = \int_0^\infty \e^{-\alpha_i y}
\Delta(y) \dd y \equiv C_i \; \;\forall \; \alpha_i >0 \;\; ({\rm for}
\; i=4,5,6)\,.
\label{infinity}
\ee
Finally, the last six integration constants are fixed by matching the
solutions at the bubble wall, and requiring that $X$ and $X'$ are
continuous at $z=0$.  This gives
\bea
0&=&\sum_{i=1}^6 A_X[\alpha_i] x_i \beta_i + A_X[\gamma_i] y_i \,,\nn \\
0&=&\sum_{i=1}^6 A_X[\alpha_i] x_i  \alpha_i\beta_i  + A_X[\gamma_i] y_i \gamma_i\,.\label{matching}
\eea

\bibliographystyle{h-physrev3} 
\bibliography{myrefs}

\end{document}